\documentclass[twocolumn, switch]{article} % Method A for two-column formatting

\usepackage{preprint}

\usepackage{enumitem}
\usepackage{amsmath, amsthm, amssymb, amsfonts}
\usepackage{bm}
\usepackage{amsmath}
\usepackage{graphicx}
\usepackage{subfigure}
\usepackage{adjustbox}
\usepackage[algoruled,algo2e]{algorithm2e}

\usepackage{setspace}
\usepackage{amssymb}
\usepackage{booktabs}
\usepackage{array}
\usepackage{color}
\usepackage{multirow}
\usepackage{multicol}
\usepackage{threeparttable}
\usepackage{url}
\usepackage[page]{appendix}
\usepackage{makecell}

\usepackage[numbers,square]{natbib}
\usepackage[utf8]{inputenc}	% allow utf-8 input
\usepackage[T1]{fontenc}	% use 8-bit T1 fonts
\usepackage{xcolor}		% colors for hyperlinks
\usepackage[colorlinks = true,
            linkcolor = blue,
            urlcolor  = blue,
            citecolor = blue,
            anchorcolor = black]{hyperref}	% Color links to references, figures, etc.
\usepackage{booktabs} 		% professional-quality tables
\usepackage{nicefrac}		% compact symbols for 1/2, etc.
\usepackage{microtype}		% microtypography
\usepackage{lineno}		% Line numbers
\usepackage{float}			% Allows for figures within multicol
\usepackage{newfloat}
\usepackage{arydshln}
\DeclareFloatingEnvironment[name={Supplementary Figure}]{suppfigure}
\usepackage{sidecap}
\sidecaptionvpos{figure}{c}

\usepackage{titlesec}
\titlespacing\section{0pt}{12pt plus 3pt minus 3pt}{1pt plus 1pt minus 1pt}
\titlespacing\subsection{0pt}{10pt plus 3pt minus 3pt}{1pt plus 1pt minus 1pt}
\titlespacing\subsubsection{0pt}{8pt plus 3pt minus 3pt}{1pt plus 1pt minus 1pt}

\usepackage{graphics}
\usepackage{hyperref}
\usepackage{color}
\usepackage{multirow}
\usepackage{hhline}
\usepackage{subfigure}
\usepackage{lipsum}
\usepackage{flushend}
\usepackage{dblfloatfix}

\title{Remote Sensing Semantic Segmentation Quality Assessment based on Vision Language Model}

\usepackage{authblk}

\author[1]{Huiying Shi}
\author[1]{Zhihong Tan}
\author[1]{Zhihan Zhang}
\author[1]{Hongchen Wei}
\author[2]{Yaosi Hu}
\author[3]{Yingxue Zhang}
\author[1]{Zhenzhong Chen}
\affil[1]{School of Remote Sensing and Information Engineering,  Wuhan University}
\affil[2]{Department of Computing, Hong Kong Polytechnic University}
\affil[3]{College of Artificial Intelligence, Tianjin University of Science and Technology}

\begin{document}

\twocolumn[ % Method A for two-column formatting
  \begin{@twocolumnfalse} % Method A for two-column formatting
  
\maketitle

\begin{abstract}

The complexity of scenes and variations in image quality result in significant variability in the performance of semantic segmentation methods of remote sensing imagery (RSI) in supervised real-world scenarios. This makes the evaluation of semantic segmentation quality in such scenarios an issue to be resolved. However, most of the existing evaluation metrics are developed based on expert-labeled object-level annotations, which are not applicable in such scenarios.
To address this issue, we propose RS-SQA, an unsupervised quality assessment model for RSI semantic segmentation based on vision language model (VLM).
Considering that segmentation performance is influenced by both model architecture and the quality of RSI, we introduce a dual-branch framework. This framework leverages a pre-trained RS VLM for semantic understanding and utilizes intermediate features from segmentation methods to extract implicit information about segmentation quality. 
Specifically, we introduce CLIP-RS, a large-scale pre-trained VLM trained with purified text to reduce textual noise and capture robust semantic information in the RS domain. Feature visualizations confirm that CLIP-RS can effectively differentiate between various levels of segmentation quality.
Semantic features and low-level segmentation features are effectively integrated through a semantic-guided approach to enhance evaluation accuracy.
To further support the development of RS semantic segmentation quality assessment, we present RS-SQED, a dedicated dataset sampled from four major RS semantic segmentation datasets and annotated with segmentation accuracy derived from the inference results of 8 representative segmentation methods.
Experimental results on the established dataset demonstrate that RS-SQA significantly outperforms state-of-the-art quality assessment models. This provides essential support for predicting segmentation accuracy and high-quality semantic segmentation interpretation, offering substantial practical value.

\end{abstract}

\vspace{0.4cm}

  \end{@twocolumnfalse} % Method A for two-column formatting
] % Method A for two-column formatting

\newcommand\blfootnote[1]{%
\begingroup
\renewcommand\thefootnote{}\footnote{#1}%
\addtocounter{footnote}{-1}%
\endgroup
}

\section{INTRODUCTION}

{\blfootnote{Corresponding author: Zhenzhong Chen, E-mail:zzchen@ieee.org}}
  
  Remote sensing imagery (RSI), distinguished by its high resolution and extensive spatial coverage, has become a cornerstone data source in various applications. Semantic segmentation, functioning as a fundamental task in object-based RSI interpretation, has been extensively implemented in diverse fields such as land cover classification \cite{7789580}, change detection, and intelligent traffic \cite{Zhang2018RiskSensAM}. Moreover, with the success of deep learning methods in computer vision, numerous deep learning-based semantic segmentation methods have emerged \cite{9258402,wang2021transformer,li2021abcnet,9487010,li2022a2,9782149,wang2022unetformer,wang2022novel,rs16162930}, demonstrating remarkable performance and evolving into potent instruments for the automated processing of RSI data.

    Downstream tasks make decisions based on the boundary and category information of the Region of Interest (ROI) obtained through RSI analysis, placing higher demands on the accuracy of semantic segmentation. However, the complexity of RSI, influenced by shooting conditions and imaging characteristics, causes segmentation precision to vary across different image instances and models. This variability significantly increases uncertainty in practical applications that lack manual annotation. 
    In order to guarantee the reliability of downstream tasks, objectively evaluating the quality of semantic segmentation results especially in the absence of labels become highly essential. 
    However, existing common semantic segmentation evaluation metrics, whether region-based \cite{ZHANG201573,article222,LIU2012144,SU2017256,article,YANG2015186} or boundary-based \cite{Cheng19052014,8863379}, rely on supervision labels and are unavailable in real-life scenarios lacking such labels.
    This raises the critical question: \textbf{“How can semantic segmentation quality be effectively assessed in an unsupervised manner?”}

\begin{figure}[!t]
      \centering
      \includegraphics[width=3.2in]{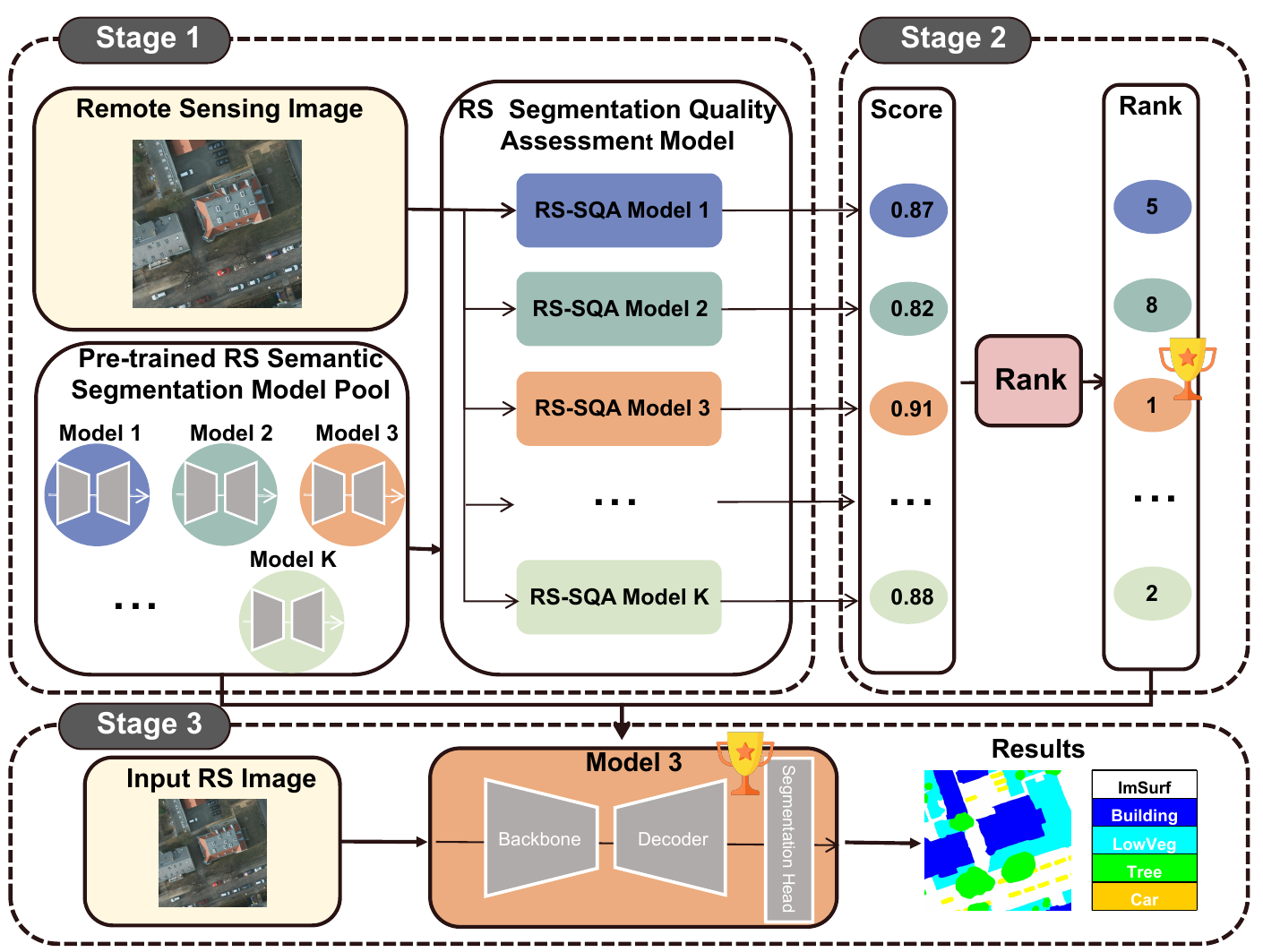}
      \caption{\textbf{The workflow for using the RS-SQA model to assist users in achieving optimal semantic segmentation.} Stage 1: Evaluate the semantic segmentation quality score for each available method. Stage 2: Rank the methods based on their quality scores and select the top-performing one. Stage 3: Apply the recommended model to segment the input image.}
      \label{figureuser}
  \end{figure}

In recent years, some methods have been devoted to solving the new problem of evaluating remote sensing semantic segmentation tasks without manual annotations. Classical methods are mainly based on the definition of human visual system perception of the ideal segmentation results \cite{Corcoran17022010}, assessing the quality from the perspective of intra-region homogeneity and inter-region heterogeneity \cite{article31666, WOODCOCK1987311, articleroclv, Zhihua}. 
However, they are more suitable for evaluating GeOBIA multi-scale segmentation results and further selecting the optimal scale parameters, but cannot truly simulate the accuracy of segmentation. Currently, Fractal \cite{chen2019effects} and Convolutional \cite{wei2021effects} have explored the simulation of the Kappa coefficient based on artificially constructed datasets. Although they provide valuable insights, they are inherently limited in capturing complex patterns of segmentation quality due to the spatial heterogeneity and spectral ambiguity of RSI. To address the challenge of RSI variability, recently, Vision-Language Models (VLMs) \cite{guo2024skysense,LHRS-Bot}, such as RemoteCLIP \cite{liu2024remoteclip}, RS-CLIP \cite{LI2023103497}, and GeoRSCLIP \cite{articlers5m}, have combined vision and language, demonstrating excellent performance in a wide range of downstream tasks.
In this paper, we propose an unsupervised semantic segmentation quality assessment method, RS-SQA, for remote sensing images based on VLM. 

 Inspired by the significant role of semantic features in the Image Quality Assessment (IQA) \cite{wang2023exploring} task, we propose that semantic features encompass multidimensional information, which is closely related to quality perception. RS-SQA leverages semantic features from CLIP-RS, a VLM that has been specifically pre-trained in the remote sensing field for geographical semantic understanding, and combines it with the intermediate layer features containing segmentation information extracted from the semantic segmentation models to form a dual-branch network. Specifically, the proposed CLIP-RS is a CLIP-based \cite{radford2021learning} model that is contrastively trained on a 10-million geographical text-image dataset. To eliminate the impact of text noise in the original data, we adopt a semantic similarity-based text purification strategy, which improves the robust semantic understanding ability of CLIP-RS.

In addition, to support the training and evaluation of the model, we establish RS-SQED, a new dataset for remote sensing semantic segmentation quality assessment. It is sampled from four commonly used remote sensing semantic segmentation datasets and labeled with the accuracy scores of eight representative deep learning-based remote sensing semantic segmentation methods. The results on the established dataset demonstrate that our method achieves comprehensive state-of-the-art (SOTA) performance, surpassing existing quality evaluation methods.

Furthermore, the experimental results on recommending the top-performance method also substantiate that RS-SQA proves its application value in facilitating the identification of the optimal semantic segmentation method. By evaluating the segmentation quality of images,it enables ranking candidate methods and recommending the most appropriate one before performing segmentation, thereby enhancing the efficiency of accurate semantic segmentation in RSIs. The workflow of using RS-SQA for RSI interpretation is shown in Fig. \ref{figureuser} .

The major contributions of our work are threefold:
\begin{enumerate}
\item{RS-SQA, a dual-branch framework, is designed for remote sensing semantic segmentation quality assessment. It can simultaneously predict semantic segmentation quality and recommend the best-performing model from a pool of eight with a 73\% accuracy.}
\item{A 10-million-scale high-quality remote sensing image-text dataset is constructed through a novel data purification strategy based on semantic similarity. Leveraging this dataset, a pre-trained remote sensing Vision Language Model CLIP-RS is proposed, enabling powerful geo-visual perception capability.}
\item{We have established the first large-scale remote sensing semantic segmentation quality evaluation dataset, RS-SQED, which covers diverse scenarios and is labeled with segmentation accuracy scores for eight different semantic segmentation methods.}
\end{enumerate}
  
  \section{Related Work} \label{sec:related_work}
  
  In this section, remote sensing semantic segmentation methods based on deep learning are reviewed. Notably, the application of the vision language model for remote sensing semantic segmentation tasks is discussed. Furthermore, the semantic segmentation quality assessment metrics are also elaborated.

\subsection{Remote Sensing Image Semantic Segmentation}\label{section RSss}
Deep learning methods have substantially improved the performance of semantic segmentation on remote sensing images. Among them, fully convolutional networks (FCNs) is a widely used architecture \cite{9076866}, enabling pixel-level spatial segmentation. To enhance the performance of FCNs on RSI, studies have incorporated techniques such as multi-scale feature fusion \cite{chen2017deeplab}, and the integration of auxiliary information (e.g., infrared images, digital surface models) \cite{kniaz2019deep}. Another widely adopted architecture is the encoder-decoder structure, exemplified by the renowned U-Net \cite{ronneberger2015u}. U-Net-like methods \cite{wang2022unetformer,rs16162930,9487010,wang2022novel} effectively combines deep and shallow features through skip connections, exhibiting excellent performance. To address the issue of easy loss of small-target information in RSI, multi-scale feature fusion-based methods have also gained traction. Representative models like FPN \cite{song2020semantic}, PSPN \cite{zhao2017pyramid}, and RefineNet \cite{lin2017refinenet} leverage the inherent pyramid structure of deep networks to combine features at different scales, preserving detailed information and achieving robust results on small-target segmentation tasks \cite{guo2020multi,cui2020sanet}.
Inspired by the success of Transformer in natural language processing, Transformer-based segmentation models have also emerged recently. These methods directly divide the image into patches, feeding them into the Transformer module for segmentation, fully exploiting Transformer's global modeling capability. Exemplary works like ResT \cite{zhang2021rest} and Segmenter \cite{strudel2021segmenter} have also demonstrated promising performance.

Recent work has embarked on exploring Vision-Language Models (VLMs) for RSI that can be fine-tuned for semantic segmentation. Prior works leverage the contrastive learning strategy of image-text pairs in RS semantic segmentation\cite{10005113}. RingMo\cite{sun2022ringmo} is the first generative self-supervised RS foundation model, pre-trained on two million RS images, achieving SOTA on four downstream tasks, including semantic segmentation. Cha et al. \cite{cha2023billion} introduce the first billion-scale foundation model in the RS field which achieves the best performance on the Potsdam and LoveDA \cite{wang2021loveda} datasets. Moreover, more remote sensing  VLMs\cite{kuckreja2024geochat,liu2024remoteclip} demonstrates robust zero-shot performance on various RS tasks, e.g., image and region captioning, visual question answering, scene classification, visually grounded conversations.
\subsection{Semantic Segmentation Quality Assessment}\label{section ml}
\subsubsection{Natural Image Semantic Segmentation Quality Assessment Methods}\label{section nml}
Segmentation quality assessment methods are mainly divided into subjective methods and objective methods depending on whether segmentation results are evaluated by human or algorithm. 
A few approaches evaluate from a subjective aspect\cite{chen2019visual}, providing human opinions to examine whether objective measures coincide with the human visual system. The objective assessment focus on measure the degree that semantic segmentation results are close to ground truth, which can be divided into region-based, contour-based and mixture metrics. Region-based metrics count the pixels correctly classified in segmentation results according to ground truth. 
The confusion matrix is an important tool for measuring the accuracy of segmentation, including four indicators: True Positive (TP), False Negative (FN), False Positive (FP) and True Negative (TN). The secondary indicators Accuracy, Precision, Recall and Specificity, as well as the tertiary indicator F1-Score, are widely applied in semantic segmentation competitions.
Contour-based metrics focus on object boundaries such as  mean distance (MD) \cite{csurka2013good} while mixture metrics consider both regions and contours simultaneously \cite{cheng2021boundary}.

\subsubsection{Remote Sensing Semantic Segmentation Quality Assessment Methods}\label{section rsml}
A wide variety of metrics for examining remote sensing image segmentation quality are proposed, generally categorized into supervised and unsupervised methods. Some supervised metrics are closely related to the objective assessment metrics discussed in Section \ref{section nml}. Others\cite{ZHANG201573,article222,LIU2012144,SU2017256,article,YANG2015186} evaluate the match between reference polygons and the computer-generated segment results, where multiple segmentation with different parameters combinations are evaluated to minimize segmentation discrepancies for further analysis. These metrics are classified into three categories: Under-Segmentation (US) metrics, Over-Segmentation (OS) metrics and Combined (UO) metrics. For unsupervised methods, some early approaches focus on measuring inter-class heterogeneity (IHE) and intra-class homogeneity (IHO) of image segments to analyze segmentation quality \cite{article31666, WOODCOCK1987311, articleroclv, Zhihua}. 
With the development of machine-learning, some researchers predict segmentation accuracy such as Kappa coefficient\cite{xia2015quality} from images. Fractal \cite{chen2019effects} extracts local morphological dimensions and global multifractal structures, while Convolutional \cite{wei2021effects} employs convolutional sparse coding to construct regressor for assessing segmentation quality. However, methods based on intra-class homogeneity struggle with the inherent dispersion in RSI. Machine learning approaches, constrained to ROI-based pixel classification, fail to generalize to deep learning methods. Overall, unsupervised quality assessment for RS semantic segmentation still requires significant advancement.
\begin{figure*}
\centering %表示居中
\includegraphics[height=8cm,width=18cm]{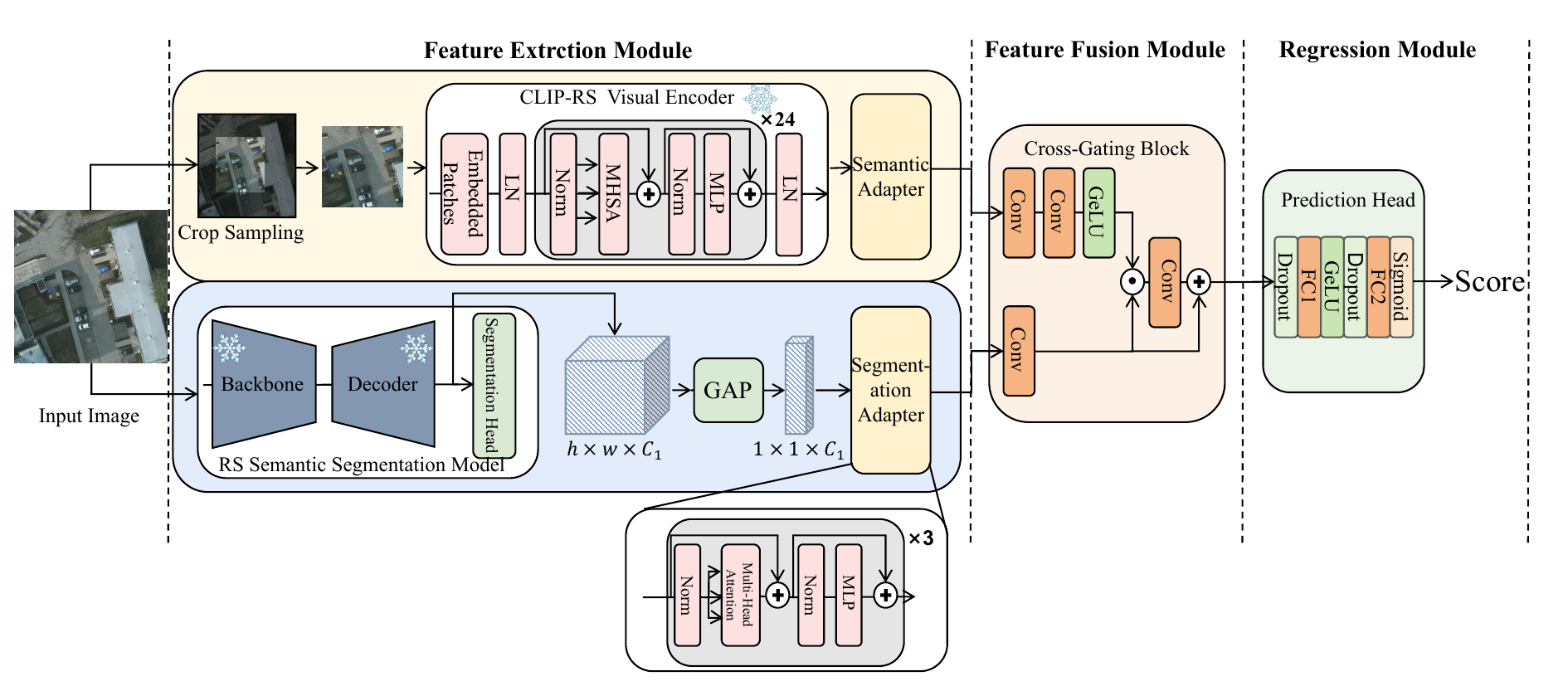}

\caption{ \textbf{Illustration of Our Framework.} High-level semantic features are extracted from CLIP-RS visual encoder, while deep segmentation features are obtained from the RS semantic segmentation model and simplified via average pooling. The features from both branches are fused using a cross-gating block and then input into a quality prediction head to generate the quality score.}
%图片的名称
\label{figure1}
%图片的标签，用于文章中的引用，注意到标签的数字与实际文章显示的数字可能不同
\end{figure*}
  
\section{Proposed Method}\label{section::method}
  
  The proposed model, RS-SQA, comprises high-level semantic feature extraction module, segmentation feature extraction module, feature fusion module, and quality prediction module. The framework of the proposed method is illustrated in Fig. \ref{figure1}. The details of each module are described in this section. 
\subsection{CLIP-RS: Vision-Language Pre-training with Data Purification for Remote Sensing}
Several studies have demonstrated the effectiveness of CLIP, a foundation model, in the IQA \cite{wang2023exploring} task due to its ability to precisely gauge subjective quality by not only general semantic information but also contextually relevant and spatially aware semantic details. Based on this, we adopt CLIP image encoder to extract semantic-aware features. 

However, to effectively transfer CLIP to the remote sensing domain while maintaining its robust semantic perception capabilities, fine-tuning on the large-scale high-quality image-text data is essential. Existing datasets are far from sufficient to meet the requirements, for instance, the data volume of RemoteCLIP is several orders of magnitude less than that of CLIP. In the light of this limitation, a large-scale, high-quality remote sensing image-text dataset comprising 10 million remote sensing images is constructed. This dataset is refined and cleaned using multimodal large language models (MLLMs) specialized in remote sensing.

The process of the construction of CLIP-RS is as follows:
\subsubsection{Data Collection}
Images are collected from globally open-source remote sensing image-text datasets. These datasets can be broadly categorized into two main types. The first type is exemplified by the RemoteCLIP dataset \cite{liu2024remoteclip}, which generates high-quality  captions via ingenious structured generation rules, containing approximately 1.5 million images. The second type consists of 8.5 million images. These images are paired with coarse semantic labels and unstructured information that might have a tenuous connection to geography. Importantly, the captions within the second type are of a heterogeneous quality. Some of them do carry a certain degree of semantic meaning and can offer assistance in understanding the related images such as `` a satellite image of landuse of forest'', while a portion is marred by noise and inaccuracies, for example, ``Google Earth to photograph by Benjamin Grant''. Despite the fact that the second type of captions enriches the data in terms of size and diversity, it also poses challenges since it causes hallucination problems in the pre-trained model \cite{zhou2024uniqa}.
\subsubsection{Data Filtering}
To effectively exploit the dataset, a novel data filtering strategy is proposed to identify captions that need to be refined from the second type captions, which is illustrated in Fig. \ref{figuredata}. 
First, a high quality semantic-aware CLIP model, denoted $\text{CLIP}_{\text{Sem}}$ is obtained through contrastive pre-training using 1.5 million high-quality texts based on the pre-trained CLIP.
Then, the similarity scores (SS) of the 8.5 million rough data are calculated by $\text{CLIP}_{\text{Sem}}$ to filter the low-quality data.
Specifically, for a given image-text pair, the SS is defined as:
\begin{equation}
SS = \frac{\mathbf{v}_I \cdot \mathbf{v}_T}{\|\mathbf{v}_I\| \|\mathbf{v}_T\|}
\end{equation}
where, $\mathbf{v}_I $ and $\mathbf{v}_T$ represent the embedding vectors extracted from visual and text spaces of $\text{CLIP}_{\text{Sem}}$, respectively. Captions are categorized into high-quality captions and low-quality captions based on SS.
\subsubsection{Data refinement}
\begin{figure*}
\centering %表示居中
\includegraphics[height=6.5cm,width=18cm]{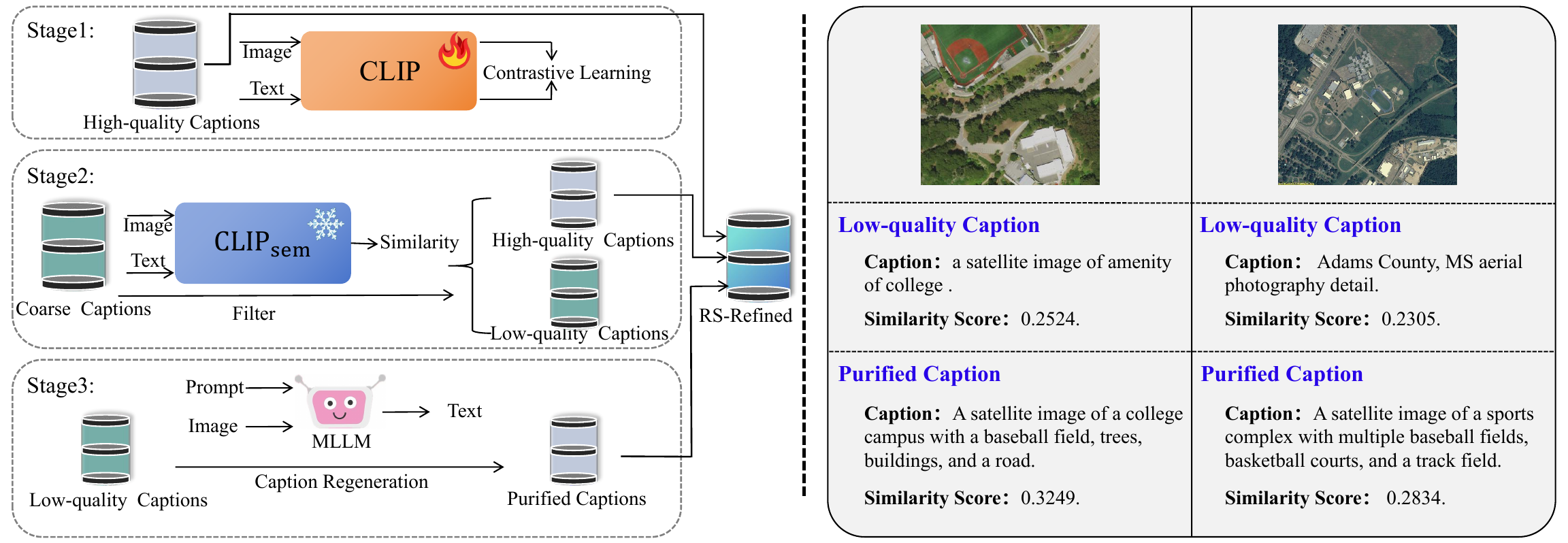}
\caption{ \textbf{Data Purification Process of the CLIP-RS Dataset.} (Left) The data purification workflow for CLIP-RS dataset. Stage 1: Train CLIP to obtain $\text{CLIP}_{\text{Sem}}$ with high-quality captions. Stage 2: Use the pre-trained $\text{CLIP}_{\text{Sem}}$ to calculate image-text similarity. Stage 3: Employ a remote sensing multi-modal large language model (MLLM) to regenerate captions for low-quality data. (Right) Examples of captioning results, showing initial low-quality image-text pairs and their corresponding purified captions. }
%图片的名称
\label{figuredata}
\end{figure*}

The grounded remote sensing VLM, GeoChat \cite{kuckreja2024geochat}, is utilized for refining captions. Due to the inherent domain disparity between remote sensing images and general images, MLLMs pre-trained on remote sensing images yield favorable results in RS visual caption tasks. To guide GeoChat in generating concise and detailed descriptions that highlight the image's key visual and contextual features, a structured input-output interaction framework was employed. The process involves three key components:
\begin{itemize}
\item{Instruction: ``Generate a brief description of the remote sensing image, highlighting key features such as the terrain, environment, layout, or other notable elements visible in the image.'' }
\item{Metacaption:``Description data of the image (insert the following data based on the actual image): [title]'' }
\item{Example: ``A satellite image of a coastal city with a network of roads, high-rise buildings, and a large harbor area.'' }
\end{itemize}
Following the input guidance, Geochat generates detailed captions for the input images.
Comparison results of low-quality captions with their corresponding refined outputs generated by GeoChat are shown in Fig. \ref{figuredata}, demonstrating the significant improvements in caption quality through refinement.
\subsubsection{Vision-Language Pre-training} Based on a diverse collection of high-quality and purified image-text pairs, the CLIP-RS is obtained by performing continual pretraining based on the CLIP model and specializing it into the remote sensing domain. This process equips the CLIP-RS with the ability of acquiring consistent semantic understanding of remote sensing images and encoding visual information.
\subsection{RS-SQA: Remote Sensing Semantic Segmentation Quality Assessment Framework}
\subsubsection{Semantic Feature Extraction Module}
 The pre-trained ViT-L-14 visual encoder from CLIP-RS is employed to capture the essential visual cues and object relationships with domain-specific prior knowledge.

Given an image $I$, it is fed into the pre-trained visual encoder to acquire the CLS tokens, which is typically considered as the high-level semantic feature vector $V_{\text{Sem}}$:
\begin{equation}
V_{\text{Sem}} = \text{Encoder}_{CLIP-RS}(I)
\end{equation}

Since the CLIP-RS visual encoder has not undergone specific training for semantic segmentation quality tasks, it is mainly related to the abstractness of quality-aware features, potentially leading to a loss of semantic information during the encoding process into latent space. Considering that fine-tuning a ViT model on the semantic segmentation quality dataset would be computationally expensive, an image adapter $A_{Sem}$ is incorporated to further interpret the semantic features into the quality-aware space. The adapter consists of three ViT blocks, and the process of extracting the semantic feature is as follows:
\begin{equation}
F_{Sem} = A_{Sem}(V_{\text{Sem}})
\end{equation}
where \( F_{Sem} \) denotes semantic features extracted from the semantic branch.
\subsubsection{Segmentation Feature Extraction Module}
Although CLIP excels at capturing global features, it primarily focuses on high-level semantic information and cannot make full use of the low-level features such as texture, blur, color, and brightness, which are crucial for pixel-level tasks. Since semantic segmentation is inherently a low-level task that relies heavily on shallow visual features, integrating segmentation-specific features can better capture these fine-grained details and provide a more effective representation of shallow spatial information relevant to segmentation.

\begin{figure*}
\centering %表示居中
\includegraphics[height=8.6cm,width=16.5cm]{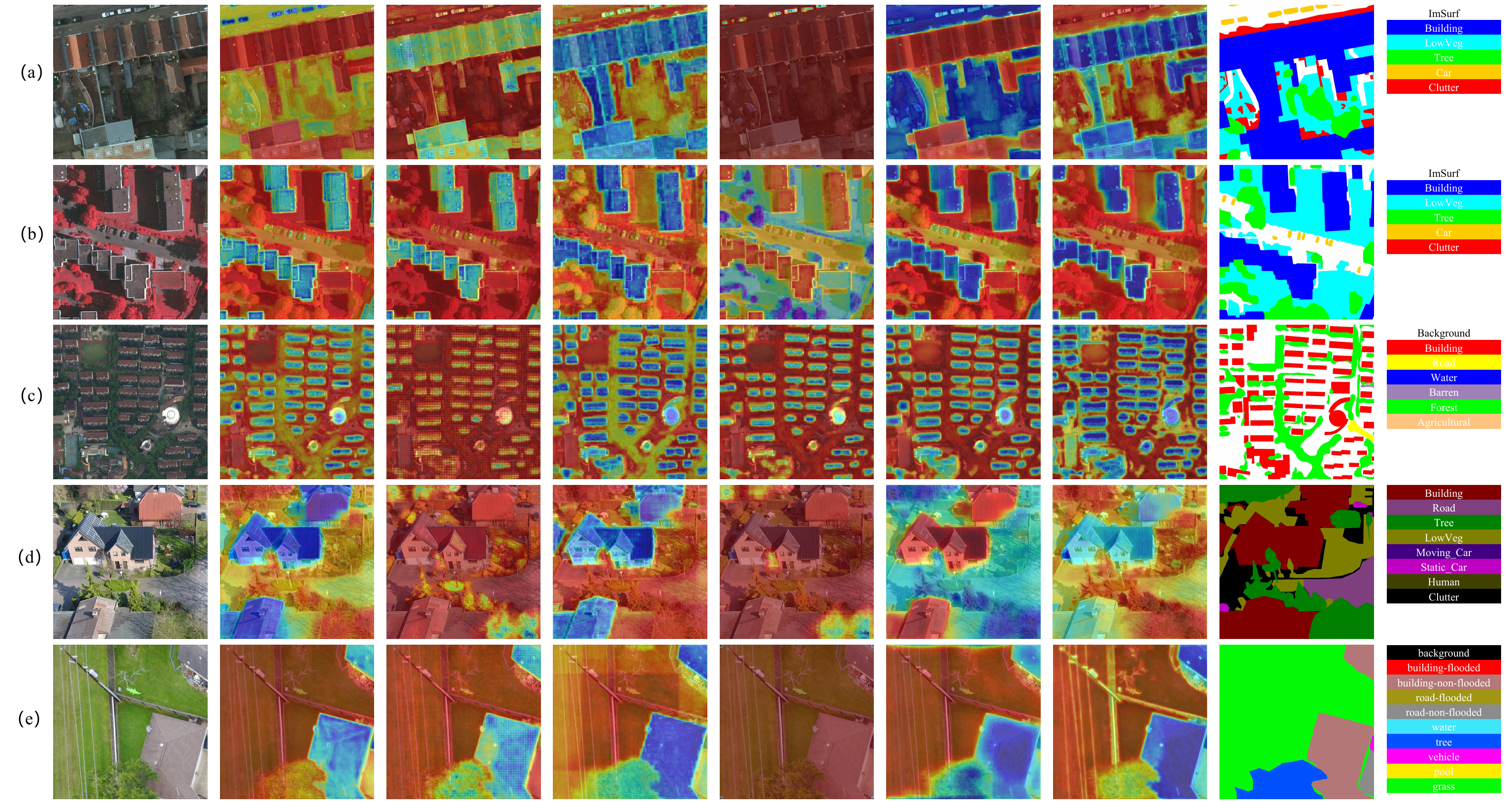}
\caption{ \textbf{Representative visualizations of features on remote sensing semantic segmentation datasets.} From left to right are raw images, the features extracted by UNetFormer \cite{wang2022unetformer}, MANet \cite{9487010}, DC-Swin \cite{wang2022novel}, BANet \cite{wang2021transformer}, A2FPN \cite{li2022a2}, and the ground truth labels, respectively. Samples from the ISPRS Potsdam, ISPRS Vaihingen, LoveDA, UAVid, and FloodNet datasets are shown in (a)-(e), respectively.}
%图片的名称
\label{fig2}
%图片的标签，用于文章中的引用，注意到标签的数字与实际文章显示的数字可能不同
\end{figure*}

Furthermore, different semantic segmentation methods employ various structures for feature extraction, leading to differences in the areas of focus within the segmentation feature maps, shown in the Fig. \ref{fig2}. These variations affect the segmentation performance across different remote sensing images. For instance, methods like UNetFormer \cite{wang2022unetformer} may prioritize capturing contextual information effectively, leading to smoother transitions between classes in the feature maps. In contrast, architectures such as MANet\cite{9487010} and DC-Swin\cite{wang2022novel} may emphasize multi-scale feature representation, which allows them to better delineate fine details in complex scenes.The observed differences in feature extraction can significantly impact segmentation performance across various remote sensing datasets. For example, in the ISPRS Potsdam dataset, the ability of BANet \cite{wang2021transformer} to capture both local textures and global structures enhances its performance, while A2FPN \cite{li2022a2} may struggle with finer details due to its focus on broader features. Consequently, the semantic segmentation features extracted from the target remote sensing segmentation method, especially from the layer preceding the classifier, are utilized. This approach enables the identification of the characteristics of each model's architecture.

Subsequently, spatial Global Average Pooling (GAP) is applied on each semantic segmentation feature map $M_{Seg}$, condensing the dimensions from  $ h \times w \times c_{1}$ to $1 \times1 \times c_{1}$ through the averaging of values within each channel.

Due to the domain gap between semantic segmentation and quality assessment, a segmentation adapter is used to bridge this issue. Given the image $I$, the process of obtaining segmentation feature is as follows:
\begin{equation}
      M_{Seg}=\text{Encoder}_{Seg}(I)
\end{equation}
\begin{equation}
      V_{Seg}=GAP(M_{Seg})
\end{equation}
\begin{equation}
      F_{Seg}=A_{Seg}(V_{Seg})
\end{equation}
where $V_{Seg}$ denotes the semantic segmentation feature vector, $A_{Seg}$ represents the segmentation adapter, and $F_{Seg}$ denotes the feature extracted from the segmentation module.
\subsubsection{Feature Fusion Module}
The CLIP-RS visual encoder demonstrates robust capabilities in representing image semantics through the vision-language contrastive pre-training. Its global semantic perspective can effectively complement the segmentation features, which are more focused on local structures. Therefore, the potential correlation between the two branches requires effective integration. Here, a feature fusion block is utilized to harness semantic feature to modulate the other branch that focuses on learning segmentation details. The novel channel-modulating block is modified from the cross-gating block\cite{tu2022maxim}, and dubbed Simple Cross-Gating Block (SCGB), for feature fusion between semantic-aware and quality-aware feature pairs.

The SCGB operates on two input tensors, $F_{Seg}$ and $F_{Sem}$, where $F_{Seg}$ originates from the segmentation branch, and $F_{Sem}$ represents features from the CLIP-based semantic branch. Through input channel projections, the projected CLIP features are fed to a gating pathway to yield the gating weights, which are then multiplied by the features from the other branch. Finally, a residual connection is adopted. The synergistic integration of semantic and segmentation features $F_{Fusion}$ is formulated as:
\begin{equation}
     W_{Sem}'= GELU(W_{Sem}F_{Sem})  
\end{equation}
\begin{equation}
 F_{Seg}'= W_{Seg}F_{Seg}
\end{equation}
\begin{equation}
F_{Fusion}=W_{Fusion}(W_{Sem}'\odot F_{Seg}')+F_{Seg}'
\end{equation}
where $W_{Sem}$, $W_{Seg}$, and $W_{Fusion}$ are learnable projection matrices for channel modulation. $GELU$ represents the activation function. $W_{Sem}'$ denotes the gating weights created by semantic features for integrating spatial semantic attention into segmentation features.
\subsubsection{Quality Regression Module}
The output of the fusion module is passed through a multi-layer perceptron (MLP) to predict quality scores, as illustrated in Fig. \ref{fig2}. The MLP consists of two fully connected (FC) layers, interleaved with a GELU activation function. To address overfitting, dropout layers with a dropout rate of 0.1 are strategically positioned between the input features and the first FC layer, as well as between the GELU activation and the second FC layer. This architecture enhances generalization by reducing reliance on individual features while balancing complexity and performance. Lastly, a sigmoid activation function is employed to map the linear outputs to values, aligning seamlessly with the requirements of segmentation quality assessment tasks ranging from 0 to 1.

\subsubsection{Loss Function}
To optimize our model, the loss function used for the proposed model consists of two parts: the Mean Squared Error (MSE) loss function and rank loss. The MSE loss measures the prediction accuracy of the model, which is defined as:
\begin{equation}
\begin{aligned}
 L_{MSE}=\frac{1}{N}\sum_{ i=1}^{N}(S_i-Q_i)^2
 \end{aligned}
\end{equation}
where $S_i$ and $Q_i$ are subjective scores and predicted scores.

To further refine the model's ability to differentiate the distribution of segmentation quality scores, inspired by \cite {kingma2013auto}, incorporating the Kullback-Leibler (KL) divergence loss function proves to be effective. This addition enhances the model's regression capability by penalizing differences between the predicted and true probability distributions. The KL divergence is defined as :
\begin{equation}
\begin{aligned}
L_{KL} = D_{KL}(P \| Q) = \sum_{i} P(i) \log \frac{P(i)}{Q(i)}
\end{aligned}
\end{equation}
where \( P \) is the true probability distribution. \( Q \) is the approximate probability distribution of quality score.
\begin{table*}[ht]
\caption{The Mean Overall Accuracy (OA) Results of the Retrained Semantic Segmentation Method on RS-SQED.}
\label{tab:model_comparison}
\centering
\resizebox{\textwidth}{!}{
\renewcommand{\arraystretch}{1.4} % Adjust row height
\begin{tabular}{ccccccccc}
% \begin{tabularx}{\textwidth}{>{\small}X>{\small}X>{\small}X>{\small}X>{\small}X>{\small}X>{\small}X>{\small}X>{\small}X}
 \hline \hline
\multirow{2}{*}{\textbf{Model}}&\multirow{2}{*}{\textbf{Time \& Venue}} & \multirow{2}{*}{\textbf{Encoder}} & \multirow{2}{*}{\textbf{Decoder}} & \multicolumn{5}{c}{\textbf{Overall Accuracy (OA)}} \\ \cline{5-9}
 & & & &\textbf{Potsdam} & \textbf{Vaihingen} & \textbf{LoveDA} & \textbf{UAVid} & \textbf{FloodNet} \\ 
 \hline
 BANet \cite{wang2021transformer}& 2021 Remote Sens. & ResT-Lite & CNN & 0.8864 & 0.8728 & 0.6715 &0.8802 
 & 0.9253 \\ 
ABCNet \cite{li2021abcnet}& 2021 ISPRS & CNN & CNN & 0.8547 & 0.8598 & 0.5131 & 0.8348 
 & 0.9020 \\ 
MANet \cite{9487010}& 2022 TGRS & ResNet50 & CNN & 0.8872 & 0.8787 & 0.6921 & 0.8859 & 0.8906 \\ 
A2FPN \cite{li2022a2}& 2022 IJRS & CNN & CNN & 0.8835 & 0.8715 & 0.6918 &0.8790 
 & 0.9157 \\ 
   UperNet(RSP-ViTAEv2-S) \cite{9782149} &2022 TGRS &RSP-ViTAEv2-S & CNN & 0.8582&  	0.8758&  	0.6865&  	0.8941&  	0.9353 \\
    UNetFormer \cite{wang2022unetformer}& 2022 ISPRS & ResNet18 & Transformer & 0.8840 & 0.8787 & 0.6713 & 0.8830 & 0.9387 \\
 DC-Swin \cite{wang2022novel}& 2022 GRSL & Swin Transformer-S & CNN & 0.8907 & 0.8802 & 0.7046 & 0.8982 
 & 0.9449 \\ 
  AerialFormer \cite{rs16162930}& 2024 Remote Sens. & Transformer & CNN & 0.8764 & 0.8760 & 0.6992 & 0.8395 
 & 0.9123 \\ 
 \hline \hline
\end{tabular}
}
\end{table*}
\renewcommand\arraystretch{1.6}
\begin{table}[ht]
\caption{Overview of the Dataset Distribution.}
\label{table1}
\centering % 让表格居中显示
\resizebox{0.5\textwidth}{!}{
\begin{tabular}{ccc>{\centering\arraybackslash}p{5.2cm}}
\hline\hline
\textbf{Database} & \textbf{Size} & \textbf{Count} & \textbf{Classes}  \\
\cline{1-4} 
ISPRS & 1024×1024 & 902  & ImSurf, Building, LowVeg, Tree, Car, Clutter   \\
LoveDA & 1024×1024 & 1669  & Background, Building, Road, Water, Barren, Forest, Agricultural   \\
UAVid & 1024×1024 & 560  & Building, Road, Tree, LowVeg, Moving-Car, Static-Car, Human, Clutter   \\
FloodNet & 1024×1024 & 3196  & Background, Building, Road, Water, Tree, Vehicle, Pool, Grass   \\
\hline\hline
\end{tabular}
}
\end{table}
\renewcommand\arraystretch{1}
The overall loss function is a weighted sum of the KL divergence loss and the MSE loss function.
\begin{equation}
\begin{aligned}
L = L_{MSE}+\alpha L_{KL}
\end{aligned}
\end{equation}
where, $\alpha$  is the arithmetic weight of $L_{KL}$ which is set to 0.5 based on \cite{wu2023exploring}. This weighted combination allows the model to balance prediction accuracy with distributional alignment, leading to improved performance.

  \section{Experimental Results} \label{sec:experiments}
  In this section, we first describe the process of constructing the remote sensing semantic segmentation quality evaluation dataset. Experimental results and analysis present the accuracy of RS-SQA in recommending the optimal semantic segmentation method.

\subsection{Database}
The RS-SQED is constructed following the three-step construction scheme outlined in \cite{chen2019effects}. The first step is to select the source images from four commonly used semantic segmentation datasets. The second step involves the selection and retraining of the semantic segmentation model. We collected eight representative RS semantic segmentation methods as candidates and retrained them on each dataset. The third step is standardization of the semantic segmentation quality score. Each of these steps is elaborated in the following sections.

\subsubsection{Image Collection}
\label{section::Image Collection}
The source images are primarily collected from four public remote sensing semantic segmentation datasets: ISPRS Vaihingen and Potsdam, LoveDA \cite{wang2021loveda}, UAVid \cite{lyu2020uavid}, and FloodNet \cite{rahnemoonfar2021floodnet}. We selected images from these datasets for two main reasons: (1) These datasets have inherently considered the diversity of semantic scenes and spatial relationship between different objects; (2) These datasets possess segmentation annotations that are generally accepted.
\begin{itemize}
\item{The ISPRS Vaihingen and Potsdam (ISPRS) dataset is released by ISPRS Commission WG II/4. The Vaihingen dataset contains 33 VFR images, with an average size of $2494\times2064$ pixels and a ground sampling distance (GSD) of 9 cm. Only the TOP image tiles are used in training and test. The Potsdam dataset contains 38 images, whose average size is 6,000 × 6,000 pixels, and the resolution is 0.5m.}
\item{The LoveDA dataset comprises 5,987 high-resolution images containing real urban and rural remote sensing images, each measuring $1024\times1024$ pixels with a GSD of 30 cm. }
\item{The UAVid dataset, as a fine-resolution Unmanned Aerial Vehicle (UAV) semantic segmentation dataset, designed for large-scale urban street scenes, consists of 300 high-resolution images of $3840\times2160$ pixels.}
\item{The FloodNet dataset is a high-resolution dataset aimed at post-flood scene understanding, comprising 1,676 images with a GSD of approximately 1.5 cm. In order to maintain consistency with the water scene content in other datasets, images with the classification category "flooded" are not used.}
\end{itemize}

We determine the sampling proportion of different datasets according to each dataset's official division. This is because the objective of semantic segmentation quality assessment is to predict segmentation quality without relying on annotations. Therefore, RS semantic segmentation models should not learn the internal patterns of RS-SQED in a supervised manner.
For ISPRS datasets, the official test tiles are used to build RS-SQED. In the case of LoveDA, where ground truth for the test set is unavailable, the validation set is used to form the RS-SQED. Similarly, for the UAVid dataset, the validation set is employed, as the ground truth for the test set was not publicly available during the construction of RS-SQED. For FloodNet, the official test set is used. The remaining portions of these datasets are used for retraining the semantic segmentation methods.

The images for quality assessment mentioned above are divided into non-overlapping patches and split into a training and testing set with an 8:2 ratio. Subsequently, they are cropped into $1024\times1024$ patches. 
The overview of the RS-SQED distribution is reported in Table \ref{table1}.

\subsubsection{Semantic Segmentation Method Training and Inference}
We explore a range of advanced methodologies for semantic segmentation in remote sensing imagery, with a primary focus on deep learning-based models. These models, which typically follow an encoder-decoder architecture, include both Transformer- and CNN-based approaches. It should be noted that our study does not include SOTA methods such as Deeplabv3+ \cite{chen2017deeplab}, PSPNet \cite{zhao2017pyramid}, ResT \cite{zhang2021rest}, and S-RS-FCN \cite{9076866}.

 We select eight methods based on diverse model structures to ensure a comprehensive evaluation. These include U-Net-like architectures (e.g., UNetFormer \cite{wang2022unetformer}, AerialFormer \cite{rs16162930}, MANet \cite{9487010}, and DC-Swin \cite{wang2022novel}), and UperNet-like frameworks (e.g., UperNet (RSP-ViTAEv2-S) \cite{9782149}). Additionally, we incorporate methods built with bilateral network frameworks (e.g., ABCNet \cite{li2021abcnet} and BANet \cite{wang2021transformer}), and Fully Convolutional Network (FCN)-like designs (e.g., A2FPN \cite{li2022a2}) to ensure diversity in model design. Table \ref{tab:model_comparison} presents a summary of different models.
 
To avoid data leakage, the remaining portions of source datasets, which do not overlap with RS-SQED, are used for retraining the semantic segmentation methods. Each method is trained following this data partition, resulting in a total of $7\times5$ models, which are subsequently used to perform semantic segmentation in RS-SQED.  
\subsubsection{Label Construction}
We adopt the overall accuracy (OA) of segmentation as the standardized metric for evaluating semantic segmentation quality. Overall accuracy (OA) is a commonly used overall indicator to measure segmentation accuracy, which denotes the percentage of correctly classified samples. Since RS-SQED is a hybrid dataset with diverse and imbalanced semantic categories, we use Overall Accuracy (OA) as a general score. The calculation of OA based on the confusion matrix\cite{forbes1995classification} is defined as:
\begin{equation}
\text{OA} = \frac{1}{N} \sum_{i=1}^{n} x_{ii} 
\end{equation}
where $n$ is the sum of columns in the confusion matrix which denotes the total number of categories. $N$ is the total number of samples in the ground truth. $x_{ii}$ is the diagonal element of the confusion matrix which represents number of correctly classified samples. 
The mean OA results of the retrained semantic segmentation method on RS-SQED are shown in Table \ref{tab:model_comparison}. It is worth noting that the OA value is computed individually for each image in the dataset.
\subsection{Experimental Setup}
\subsubsection{Implementation Details}

We train CLIP-RS on a single-node 7×NVIDIA GeForce RTX 4060 Ti machine. We initialize the CLIP/ViT-L-14 with the OpenAI model weights based on the performance of the initial model weights. The learning rate is set to 3e-9, and the corresponding batch size is set to $7\times16$.

 We train RS-SQA on an NVIDIA TITAN XP machine. The AdamW optimizer is employed to optimize the model parameters, with the learning rate set to 1e-4. To ensure effective training, a two-stage learning rate scheduling strategy is adopted: lr\_scheduler.LinearLR for warm-up and lr\_scheduler.StepLR for formal training.

We retrain the candidate models on an NVIDIA TITAN XP machine. The semantic segmentation model's training procedure adheres to the learning rate and optimizer configurations specified in the open-source code. As we are not comparing different seg methods to discuss their performance, we set the batch size to 2 during training, such that the results from different methods could share the same configuration.

\subsubsection{Evaluation Metrics}
\begin{figure*}
\centering %表示居中
\includegraphics[height=17cm,width=16cm]{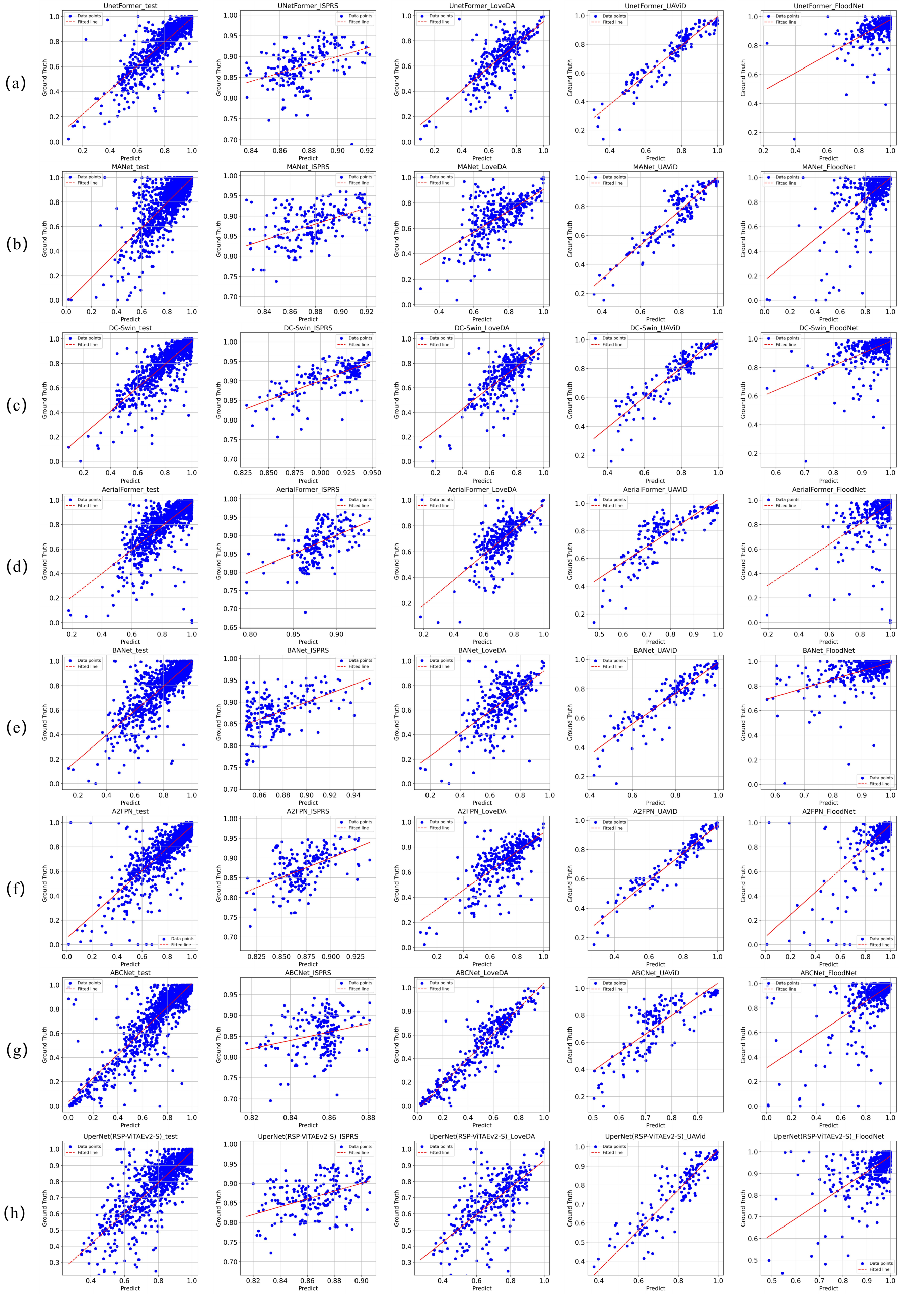}
\caption{ \textbf{Scatter plots between the predicted Overall Accuracy (OA) and the ground truth OA.} The predicted OA is derived from models trained on ground truth segmented by UNetFormer \cite {wang2022unetformer}, MANet \cite{9487010}, DC-Swin \cite{wang2022novel}, AerialFormer \cite{rs16162930}, BANet \cite{wang2021transformer}, A2FPN \cite{li2022a2}, ABCNet \cite{li2021abcnet}, and UperNet(RSP-ViTAEv2-S) \cite{9782149}, respectively (corresponding to subplots a, b, c, d, e, f, g and h). From left to right, the results correspond to RS-SQED, ISPRS Vaihingen and Potsdam, LoveDA, UAVid, and FloodNet datasets.}
%图片的名称
\label{fig-performanceour}
%图片的标签，用于文章中的引用，注意到标签的数字与实际文章显示的数字可能不同
\end{figure*}
%%%%%%%%%%%%%%%%%%%%%%%%%%%%%%%%%%%%%%%%%%%%%%%%%%%%%%%%%%%%%%%%%%%%%%%%%%%%%%%%%
\begin{table*}[h!]
    \centering
    \caption{Performance Comparison of the Related Models on RS-SQED. The Bold Results and the Underlined Results Indicate the Top and the Second Performers.}
    \label{table::perormanc1}
    \setlength{\tabcolsep}{2pt}
    \resizebox{\textwidth}{!}{
    % \begin{tabular}{c|c|c:c:c:c|c:c|c:c|c:c|c:c|c:c}
    \renewcommand{\arraystretch}{1}
    \begin{tabular}{cccccccccccccccccccc}
    \hline
    \hline
    % \cline{1-18}
    \multirow{2}{*}{\small\textbf{Seg Method}}   & \small\multirow{2}{*}{\textbf{Model}} 
    & \multicolumn{4}{c}{\small \textbf{RS-SQED}} &  & \multicolumn{2}{c}{\textbf{\small ISPRS}} & &  \multicolumn{2}{c}{\small \textbf{LoveDA}} & & \multicolumn{2}{c}{\textbf{\small UAVid}} &&   \multicolumn{2}{c}{\textbf{\small FloodNet}} \\ 
    \cline{3-6}  \cline{8-9} \cline{11-12} \cline{14-15} \cline{17-18} 
    & & \small \textbf{PLCC }& \small \textbf{SROCC} & \small \textbf{RMSE} & \small \textbf{KROCC} &  &\small \textbf{PLCC }& \small \textbf{SROCC} & & \small \textbf{PLCC }&\small  \textbf{SROCC} & & \small \textbf{PLCC }& \small \textbf{SROCC} &  &\small \textbf{PLCC }& \small \textbf{SROCC}  \\ 
   \cline{1-18}

    \multirow{7}{*} {\makecell{\small UperNet\\(RSP-ViTAEv2-S)}}
       &\small BLIINDS-II  \cite{6172573}  &\small 0.1463 	 &\small 0.1809 	 &\small 0.1675 	 &\small 0.1207   &  &\small 0.2348 
 
 &\small  0.2248 
  &   &\small 	0.2023 
  &\small 	0.2155 &    &\small 	0.4949 	 &\small -0.4671  &   &\small 	0.1955  &\small 	0.2425 
 \\
        &\small BRISQUE  \cite{mittal2012no}  &\small 0.3799  &\small 	0.4333  &\small 	0.1530 	 &\small 0.3072  &\small 	  &\small 0.1706 
 
  &\small 	0.0558 

  &\small 	  &\small 0.3310 &\small  	0.0990 &\small    &\small 	0.2823 &\small  	0.1828 &\small    &\small 	0.2152 &\small  	0.2314 
\\
         &\small  DEIQT \cite{qin2023data}   &\small  \underline{0.7826} 	 &\small\underline{0.8333} &\small 	\underline{0.3607} &\small  	0.2152 
  &\small   &\small 	0.3781 
 &\small 0.3674 
 &\small   &\small 	0.6139  &\small 	0.5939  &\small   &\small 	\underline{0.8115 } &\small 	0.7556  &\small   &\small 	0.4600 	 &\small 0.6344  
\\
         &\small  HyperIQA \cite{su2020blindly}   &\small 0.6629  &\small 	0.7320 &\small  	0.1282  &\small 	0.5518 &\small    &\small 	0.3756 
 	 &\small \underline{0.3844}
  &\small 	 &\small 0.4299  &\small 	0.4370 &\small  	  &\small 0.2022  &\small 	0.1223  &\small 	  &\small 0.3614 &\small  	0.5337 
\\
 &\small  MANIQA \cite{yang2022maniqa}  &\small 0.7207 &\small  	0.7872  &\small 	0.1182  &\small 	0.5969 &\small    &\small 	0.2325 

 &\small  	0.1581 
 
 	 &\small   &\small 0.3776 	 &\small 0.3667  &\small   &\small 	0.6875  &\small 	0.6688 &\small   &\small  	0.3679  &\small 	0.4967 
\\
         &\small  Fractal \cite{chen2019effects}  &\small  0.6914  &\small  	0.8186 	 &\small  0.1346  &\small  	\underline{0.6601}  &\small   &\small  	\textbf{0.5849} 	 &\small  \textbf{0.5936}  &\small   &\small  	\underline{0.6621}  &\small  	\textbf{0.7072} &\small   &\small   	0.7566  &\small  	\underline{0.8465} &\small     &\small 	\underline{0.6489} 	 &\small  \underline{0.7471}
\\
% \dashline{2-4}
         &\small  Ours  &\small  \textbf{0.8553}  &\small 	\textbf{0.8765}  &\small 	\textbf{0.0884}  &\small 	\textbf{0.7131}  &\small  &\small 	\underline{0.4200}  &\small 	0.3789 
 
 	 &\small  &\small \textbf{0.7059}   &\small \underline{0.6853 }	 &\small  &\small \textbf{0.8878}  &\small 	\textbf{0.8921} 	 &\small   &\small \textbf{0.7205}  &\small 	\textbf{0.7869}
\\
        \midrule

       \multirow{7}{*} { \small UNetFormer}
        &\small BLIINDS-II  \cite{6172573}  &\small 0.1686 &\small  	0.1914  &\small 	0.1735  &\small 	0.1274  &\small  &\small 	0.2507 
  &\small 	0.2023  
 	 &\small  &\small 0.1293  &\small 	0.2135 &\small   &\small 	0.5055 &\small 	-0.4746  &\small  &\small 	0.1891  &\small 	0.2207 \\

        &\small BRISQUE  \cite{mittal2012no}  &\small 0.4047  &\small 0.4633  &\small 0.1556  &\small 0.3323  &\small  &\small 0.2809 
  &\small 0.2288 
  &\small  &\small 0.3444  &\small 0.0792  &\small  &\small 0.3025  &\small 0.2109  &\small  &\small 0.2362  &\small 0.2572 
\\
         &\small  DEIQT \cite{qin2023data}   &\small  \underline{0.7720}  &\small  0.8177  &\small  \underline{0.0982} &\small  \underline{0.6711} &\small   &\small 0.3781 
  &\small0.3674 
 &\small   &\small \underline{0.6890}  &\small 0.6219  &\small   &\small \underline{0.7638}  &\small 0.7633  &\small  &\small  0.4693  &\small 0.6276 
\\
         &\small  HyperIQA \cite{su2020blindly}   &\small  0.7344  &\small  0.7877   &\small  0.1158   &\small  0.6034   &\small  &\small  0.3799 
   &\small  0.4431 
   &\small  &\small  0.5498  &\small 0.4637   &\small  &\small  0.1055  &\small  0.0955 &\small   &\small 0.3389   &\small 0.4550 
\\
 &\small  MANIQA \cite{yang2022maniqa}  &\small 0.7653  &\small 0.8101  &\small 0.1117  &\small 0.6176 &\small   &\small 0.2918 
  &\small 0.0982 
  &\small  &\small 0.4962  &\small 0.3721  &\small  &\small 0.5668 &\small 0.5575 &\small  &\small 0.3288  &\small 0.4977 
\\
         &\small  Fractal \cite{chen2019effects}  &\small  0.6938   &\small  \underline{0.8316 }  &\small  0.1387   &\small  0.6709   &\small  &\small  \textbf{0.6902}   &\small  \textbf{0.6341}  &\small  &\small  0.6524  &\small  \textbf{0.7013} &\small   &\small  0.7007  &\small  \underline{0.8036}  &\small   &\small  \textbf{0.6643}   &\small  \underline{0.7660 }
\\
%\hdashline
         &\small  Ours  &\small   \textbf{0.8807} &\small  	\textbf{0.9027}  &\small 	\textbf{0.0806}  &\small 	\textbf{0.7366}  &\small  &\small 	\underline{0.6415}  &\small 	\underline{0.6250}  &\small  &\small 	\textbf{0.7342} 	 &\small \underline{0.6963}  &\small  &\small 	\textbf{0.9382} 	 &\small \textbf{0.9401} &\small  &\small  	\underline{0.6556}  &\small 	\textbf{0.7873} 

\\
        \midrule

        \multirow{7}{*} { \small MANet}
          &\small BLIINDS-II  \cite{6172573} &\small 0.0977 &\small  	0.1700 	 &\small 0.1906  &\small 	0.1131  &\small  &\small 	0.2175 
 
  &\small 	0.1215 
 &\small   &\small 	0.0172  &\small 	0.2072  &\small  &\small 	0.5262  &\small 	-0.5004  &\small  &\small 	0.1252  &\small 	0.1574 \\

         &\small BRISQUE  \cite{mittal2012no}  &\small 0.3134  &\small 0.3807  &\small 0.1785  &\small 0.2703  &\small  &\small0.2871 
  &\small 0.2141 
  &\small  &\small 0.1149  &\small 0.0589  &\small  &\small 0.2950  &\small 0.2243  &\small  &\small 0.1601  &\small 0.1842 
\\
         &\small   DEIQT \cite{qin2023data}             &\small  0.6445  &\small  0.6991  &\small  \underline{0.1260 }
  &\small 0.5816 
 &\small  &\small 0.3764 
  &\small 0.3796 
 &\small   &\small  \textbf{0.6802}  &\small  0.6000  &\small   &\small \underline{0.7684 } &\small  0.7282  &\small  &\small  0.5394  &\small  0.6052 \\
         &\small  HyperIQA  \cite{su2020blindly}               &\small  0.6273  &\small  0.7038  &\small 0.1391 
 &\small 0.5371 
 &\small   &\small 0.4130  
  &\small  \underline{0.4313}  &\small  &\small  0.5937  &\small  0.5306  &\small  &\small  0.1774  &\small  0.0572  &\small  &\small  0.5041  &\small  0.5632 \\
 &\small  MANIQA\cite{yang2022maniqa}  &\small 0.6092  &\small 0.6895  &\small 0.1503  &\small 0.5156  &\small  &\small 0.2432   &\small0.0711 
  &\small  &\small 0.4954  &\small 0.4316  &\small  &\small 0.4661 &\small 0.4211 &\small   &\small 0.3798  &\small 0.5134 
\\
         &\small  Fractal \cite{chen2019effects}                                  &\small  \underline{0.6758}  &\small  \underline{0.7435}  &\small  0.1559 
 &\small  \textbf{0.6576 }
 &\small  &\small  \textbf{0.5288}  &\small  \textbf{0.4827} &\small   &\small  0.6422  &\small \textbf{0.6940}  &\small  &\small  0.7062  &\small  \underline{0.7968}  &\small  &\small  \underline{0.6496}  &\small  \textbf{0.7673 }\\
        %\hdashline
         &\small  Ours                                    &\small  \textbf{0.7968}  &\small 	\textbf{0.8126}  &\small 	\textbf{0.1088}  &\small 	\underline{0.6374}  &\small  &\small 	\underline{0.4240}  &\small 	0.3614 	 &\small  &\small \underline{0.6688}  &\small 	\underline{0.6361}  &\small  &\small 	\textbf{0.9135} &\small  	\textbf{0.8297} 	 &\small  &\small \textbf{0.6769}  &\small 	\underline{0.6972} 
 \\ 
        \midrule

       \multirow{7}{*} { \small AerialFormer}
         &\small BLIINDS-II  \cite{6172573} &\small 0.0888 	 &\small 0.1419  &\small 	0.1868  &\small 	0.0927  &\small  &\small 	0.2034 
  &\small 	0.1590 
  &\small  &\small 	0.0936  &\small 	0.1702  &\small  &\small 	0.4968 &\small  	-0.4722  &\small  &\small 	0.0940  &\small 	0.1356 \\

        &\small BRISQUE  \cite{mittal2012no}  &\small 0.3393  &\small 0.4023  &\small 0.1723  &\small 0.2837  &\small  &\small 0.1977 
  &\small0.1129 
  &\small  &\small 
0.3338  &\small 	0.0513  &\small  &\small 	0.2974  &\small 	0.2045 &\small  &\small  	0.1913 	 &\small 0.2269 
\\
         &\small   DEIQT \cite{qin2023data}                      &\small  0.6560  &\small  0.7259  &\small  \underline{0.1163 }
 &\small 0.5944 
 &\small   &\small 0.3363 
  &\small  0.3319  &\small   &\small 0.6063 
  &\small  0.5588  &\small  &\small  \underline{0.7451}  &\small  0.7228  &\small  &\small  0.4486  &\small  0.5316 \\ 
         &\small   HyperIQA \cite{su2020blindly}                     &\small  0.6298  &\small  0.6935  &\small 0.1295 
 &\small 0.5278 
 &\small  &\small 0.1983 
  &\small  0.1778 
 &\small   &\small  0.5200  &\small  0.5275  &\small  &\small  0.5138  &\small  0.5830  &\small  &\small  0.3220  &\small  0.4449 \\ 
 &\small  MANIQA \cite{yang2022maniqa}  &\small 0.6449  &\small 	0.7389  &\small 	0.1293 &\small  	0.5596  &\small  &\small 	0.2818 
  &\small 	0.2358 
  &\small  &\small 	0.3920  &\small 	0.3347 &\small   &\small 	0.7021 &\small  	0.6771  &\small  &\small 	0.3084  &\small 	0.4994 \\
         &\small  Fractal  \cite{chen2019effects}                                  &\small  \underline{0.6826}  &\small  \underline{0.7474}  &\small  0.1516 
 &\small  \underline{0.6626 }
 &\small   &\small \textbf{0.6049 
}  &\small  \textbf{0.6424 
}  &\small  &\small  \underline{0.6751}  &\small  \textbf{0.7273 } &\small   &\small 0.7001  &\small  \underline{0.8065}  &\small  &\small  \textbf{0.6557 }  &\small  \underline{0.7652} \\
        %\hdashline
         &\small  Ours                                  &\small \textbf{0.8148} &\small  \textbf{	0.8725}  &\small 	\textbf{0.0973} &\small  	\textbf{0.7052}  &\small 	 &\small  \underline{0.5180 
}  &\small 	 \underline{0.5741 
}  &\small 	 &\small \textbf{0.6990} &\small  	 \underline{0.6699} 	 &\small  &\small \textbf{0.8931} 	 &\small \textbf{0.8662}  &\small  &\small 	 \underline{0.6351} 	 &\small \textbf{0.7993}
 \\ 
        \midrule

       \multirow{7}{*} {\small DC-Swin}
         &\small BLIINDS-II  \cite{6172573} &\small 0.1922  &\small 	0.2223  &\small 	0.1583  &\small 	0.1495  &\small  &\small 	0.0755 
  &\small 	0.0627 
  &\small  &\small 	0.1497 &\small  	0.2170  &\small  &\small 	0.5052  &\small 	-0.4781 	 &\small  &\small 0.2064  &\small 	0.2411 \\

        &\small BRISQUE  \cite{mittal2012no}  &\small 0.4010 &\small  	0.4441  &\small 	0.1432  &\small 	0.3171  &\small 	 &\small 0.5071 
  &\small 	0.4604 
 &\small 	 &\small 0.0742  &\small 	0.0485  &\small  &\small 	0.3513  &\small 	0.2664  &\small  &\small 	0.2120  &\small 	0.2286 
\\
         &\small  DEIQT  \cite{qin2023data}                  &\small  \underline{0.7664}  &\small  \underline{0.8112 } &\small  \underline{0.0946 }
 &\small 0.6636 
 &\small   &\small 0.6325 
  &\small  0.5708 
  &\small   &\small0.6445  &\small  0.5512  &\small   &\small \underline{0.8123}  &\small  0.7730  &\small  &\small  0.4763  &\small  0.6544 \\ 
         &\small  HyperIQA  \cite{su2020blindly}                     &\small  0.6944  &\small  0.7544  &\small 0.1165 
 &\small 0.5710 
 &\small  &\small  0.6268 
  &\small  0.4927 
  &\small  &\small  0.4594  &\small  0.4265  &\small  &\small  0.1449  &\small  0.0845  &\small  &\small  0.3562  &\small  0.3880 \\
 &\small  MANIQA  \cite{yang2022maniqa}  &\small 0.7425 	 &\small 0.8013  &\small 	0.1092  &\small 	0.6075 &\small   &\small 	0.5491 
  &\small 	0.5092 
  &\small  &\small 	0.4588  &\small 	0.4002  &\small  &\small 	0.6748  &\small 	0.6051  &\small  &\small 	0.3482  &\small 	0.5162 
\\
         &\small  Fractal  \cite{chen2019effects}                                  &\small  0.7156  &\small  0.7506  &\small 0.1267 
  &\small  \underline{0.6826} 
 &\small  &\small   \underline{0.7099 
}  &\small  \textbf{0.6814 
}  &\small  &\small  \underline{0.6532}  &\small   \textbf{0.7418}  &\small   &\small 0.6759 &\small  \underline{0.7810}  &\small  &\small   \textbf{0.6729}  &\small  \underline{0.7748} \\ 
        %\hdashline
         &\small  Ours                                  &\small  \textbf{0.8678}  &\small 	\textbf{0.9063} &\small  	\textbf{0.0802}  &\small 	\textbf{0.7411}  &\small  &\small 	\textbf{0.7151} &\small  	\underline{0.6700}  &\small  &\small 	\textbf{0.7125} 	 &\small \underline{0.6556} 	 &\small  &\small \textbf{0.9068}  &\small 	\textbf{0.9004}  &\small 	 &\small \underline{0.5965}  &\small 	\textbf{0.8284} 
 \\ 
        \midrule

      \multirow{7}{*} {\small BANet}
        &\small BLIINDS-II  \cite{6172573} &\small 0.0958  &\small 	0.1584  &\small 	0.1945  &\small 	0.1053 &\small  &\small  	0.2928 
  &\small 	0.2330 
 &\small  &\small  	0.0332  &\small 	0.1881  &\small  &\small 	0.4988  &\small 	-0.4614  &\small  &\small 	0.1070  &\small 	0.1789 \\

       &\small BRISQUE  \cite{mittal2012no}  &\small 0.3460 	 &\small 0.4239  &\small 	0.1788 	 &\small 0.3027  &\small  &\small 	0.2024 
 	 &\small 0.1509 
 	 &\small  &\small 0.3837  &\small 	0.0920  &\small  &\small 	0.2755 &\small  	0.2068 &\small   &\small 	0.1746  &\small 	0.2126 
\\
         &\small  DEIQT  \cite{qin2023data}                    &\small  0.6829  &\small  0.7554  &\small  0.1871 
 &\small 0.6235 
 &\small  &\small 0.3611 
  &\small  0.3075 
  &\small  &\small  0.4741  &\small  0.4348  &\small  &\small   \underline{0.7919 } &\small  0.7615  &\small   &\small 0.3210  &\small  0.5220 \\ 
         &\small  HyperIQA   \cite{su2020blindly}                  &\small  0.6848  &\small  0.7079  &\small 0.1274 
 &\small 0.5328 
 &\small  &\small  0.4246 
  &\small  0.4639 
  &\small  &\small  0.5354  &\small  0.5109  &\small  &\small  0.4881  &\small  -0.2050  &\small   &\small 0.2896  &\small  0.3957 \\
 &\small  MANIQA \cite{yang2022maniqa}  &\small  \underline{ 0.7296}  &\small 	 \underline{0.7731}  &\small 	 \underline{0.1236}  &\small 	0.5838 &\small   &\small 	0.2356 
 &\small  	0.1644 
  &\small  &\small 	0.4366  &\small 	0.3547  &\small 	 &\small 0.6580  &\small 	0.5519  &\small  &\small 	0.2870  &\small 	0.4230 
\\
         &\small  Fractal  \cite{chen2019effects}                          &\small  0.6631  &\small  0.7417  &\small 0.1602 
 &\small   \underline{0.6600}
 &\small  &\small \textbf{0.6915 
}  &\small  \textbf{0.6475 
}  &\small  &\small  \textbf{0.6737}  &\small  \textbf{0.7114}
&\small   &\small  0.6931  &\small   \underline{0.7747 } &\small  &\small  \underline{0.6446 } &\small  \textbf{0.7572} \\
        %\hdashline
         &\small  Ours                              &\small  \textbf{0.8468} 	 &\small \textbf{0.8737}  &\small 	\textbf{0.0931} 	 &\small \textbf{0.7020}  &\small  &\small 	\underline{0.5428}  &\small 	\underline{0.4475} &\small   &\small \underline{0.6292}  &\small 	\underline{0.5927} &\small 	 &\small \textbf{0.9149}  &\small 	\textbf{0.9104} &\small   &\small 	\textbf{0.6833}  &\small 	\underline{0.7342}
\\
        \midrule

       \multirow{7}{*} {\small A2FPN}
         &\small BLIINDS-II  \cite{6172573} &\small 0.1188  &\small 	0.1703 &\small  	0.1887  &\small 	0.1131  &\small 	 &\small 0.1971 
  &\small 0.1489 
 &\small   &\small 	0.0353  &\small 	0.1953  &\small  &\small 	0.5112 &\small  	-0.4717  &\small 	 &\small 0.1371  &\small 	0.1952 \\

        &\small BRISQUE  \cite{mittal2012no}  &\small 0.3083  &\small 	0.4071  &\small 	0.1756  &\small 	0.2911 	 &\small  &\small 0.3073 
 	 &\small 0.2337 
  &\small  &\small 	0.0809  &\small 	0.0844  &\small  &\small 	0.3008 	 &\small 0.2089  &\small  &\small 	0.1848 	 &\small 0.2115 
\\
         &\small  DEIQT \cite{qin2023data}                     &\small   \underline{0.7125}  &\small   \underline{0.7738}  &\small  0.1780 
 &\small 0.6436 
 &\small  &\small  0.4270 
  &\small 0.3424 
  &\small   &\small  \underline{0.6541}  &\small  0.6375  &\small  &\small   \underline{0.8197}  &\small  0.7953  &\small  &\small  0.4703  &\small  0.5540 \\ 
         &\small  HyperIQA \cite{su2020blindly}                      &\small  0.5964  &\small  0.6966  &\small 0.1480 
 &\small 0.5243  &\small 
 &\small  0.3164 
  &\small  0.2552 
  &\small  &\small  0.4391  &\small  0.4463  &\small  &\small  0.4900  &\small  -0.2453  &\small   &\small 0.3835  &\small  0.4167 \\ 
 &\small  MANIQA \cite{yang2022maniqa}  &\small 0.6425  &\small 	0.7591  &\small 	 \underline{0.1413}  &\small 	0.5779  &\small  &\small 0.3456 
  &\small 	0.2697 
  &\small  &\small 	0.5583 &\small  	0.4474 &\small   &\small 	0.6356  &\small 	0.6159  &\small 	 &\small 0.3400  &\small 	0.5093 
\\
         &\small  Fractal \cite{chen2019effects}                           &\small  0.6608  &\small  0.7501  &\small 0.1550 
 &\small   \underline{0.6684 }
 &\small   &\small \textbf{0.5509}  &\small \textbf{0.6005}  &\small  &\small  0.6538  &\small \underline{0.7262} &\small   &\small 0.7249  &\small   \underline{0.8065}  &\small   &\small \underline{ 0.6515}  &\small   \underline{0.7672} \\
        %\hdashline
         &\small  Ours                                &\small \textbf{0.8569} 	 &\small \textbf{0.8940}  &\small 	\textbf{0.0948} 	 &\small \textbf{0.7338} 	 &\small  &\small  \underline{0.4922 
}  &\small 	 \underline{0.4640 
} 	 &\small  &\small \textbf{0.7819}  &\small 	\textbf{0.7324}	 &\small  &\small \textbf{0.9439}  &\small 	\textbf{0.9341} 	 &\small  &\small \textbf{0.7462}  &\small 	\textbf{0.8077}
 \\ 
        \midrule

       \multirow{7}{*} {\small ABCNet}
         &\small BLIINDS-II  \cite{6172573} &\small 0.1804 	 &\small 0.1991  &\small 	0.2635  &\small 	0.1318  &\small  &\small 	0.1560 
  &\small 	0.1230 
  &\small  &\small 	0.0309  &\small 	0.1795  &\small  &\small 	0.4941  &\small 	-0.4711  &\small 	 &\small 0.1735  &\small 	0.1978 \\
        &\small BRISQUE  \cite{mittal2012no}  &\small 0.3664 	 &\small 0.4074  &\small 	0.2402 	 &\small 0.0449 
 	 &\small  &\small 0.1477 
  &\small 	0.0449 
  &\small  &\small 	0.2809  &\small 	0.1196 &\small  &\small  	0.2513  &\small 	0.1778  &\small  &\small 	0.1864  &\small 	0.1916 
\\
         &\small  DEIQT  \cite{qin2023data}                      &\small \underline{ 0.8351}  &\small  \underline{0.8070}  &\small  \underline{0.1212}  &\small \underline{ 0.6796}  &\small  &\small 0.3011 
 
  &\small 0.1880 
 
  &\small  &\small  \underline{0.8869}  &\small  \underline{0.8678}  &\small  &\small  \underline{0.7557}  &\small  0.6987  &\small  &\small 0.5971 
 &\small 0.6430 
\\ 
         &\small  HyperIQA   \cite{su2020blindly}                &\small  0.7024  &\small  0.7779  &\small  0.1873  &\small  0.6028  &\small  &\small  0.2439 
  &\small  0.1764 
  &\small  &\small  0.4257  &\small  0.4914  &\small  &\small  0.2697  &\small  -0.1931  &\small  &\small 0.3259 
 &\small 0.5894 
\\  
 &\small  MANIQA  \cite{yang2022maniqa}  &\small 0.8037 	 &\small 0.7892 &\small  	0.1540  &\small 	0.6140 &\small  &\small  	0.1373 
 &\small  	0.0892 
 &\small   &\small 	0.7799  &\small 	0.7579  &\small  &\small 	0.5419  &\small 	0.4748  &\small  &\small 	0.4329  &\small 	0.5514 
\\
         &\small  Fractal  \cite{chen2019effects}                                 &\small  0.6708  &\small  0.8052  &\small  0.2272  &\small  0.6402  &\small   &\small \textbf{0.4636 
}  &\small \textbf{0.5014 
}  &\small  &\small  0.6689  &\small  0.6757  &\small  &\small  0.7284  &\small  \underline{0.8079 } &\small  &\small  \underline{0.6522 }
 &\small  \underline{0.7281 }
\\ 
        %\hdashline
         &\small  Ours                             &\small \textbf{0.9021} 	 &\small \textbf{0.8963}  &\small 	\textbf{0.1096}  &\small 	\textbf{0.7430} &\small   &\small 	\underline{0.4212 
}  &\small 	\underline{0.3785 
}  &\small  &\small 	\textbf{0.9340}  &\small 	\textbf{0.9096}  &\small  &\small 	\textbf{0.8576} &\small  	\textbf{0.8102}  &\small  &\small 	\textbf{0.6875}  &\small 	\textbf{0.7584} 

\\ 
    \hline\hline
\end{tabular}
}
\end{table*}
\renewcommand\arraystretch{1}
%%%%%%%%%%%%%%%%%%%%%%%%%%%%%%%%%%%%%%%%%%%%%%%%%%%%%%%%%%%%%%%%%%%%%%%%%%%%%%%%%
\renewcommand\arraystretch{1.8}
\begin{table*}[ht]
   \caption{Precision Comparison of the Best Method Prediction among Eight Models. The Bold Indicates the Top One Performer.}
   \label{table7}
   \centering
 \resizebox{\textwidth}{!}{ % 调整表格宽度为页面一半
   \begin{tabular}{cccccccccccccccc}
   \hline\hline
   \multirow{2}{*}{ \large \textbf{Model}} & \multirow{2}{*}{ \large \textbf{Time \& Venue}} & \multicolumn{2}{c}{ \large \textbf{RS-SQED}} && \multicolumn{2}{c}{ \large \textbf{ISPRS}} & & \multicolumn{2}{c}{ \large \textbf{LoveDA}} && \multicolumn{2}{c}{\large \textbf{UAVid}} & &\multicolumn{2}{c}{ \large \textbf{FloodNet}} \\
   \cline{3-4} \cline{6-7}\cline{9-10}\cline{12-13}\cline{15-16}
   & &  \large \textbf{P@1 } & \large  \textbf{P@3} & &  \large \textbf{P@1 } &  \large \textbf{P@3}& &   \large \textbf{P@1 } & \large  \textbf{P@3} & & \large  \textbf{P@1 } & \large  \textbf{P@3} &&  \large  \textbf{P@1 } &  \large \textbf{P@3} \\
   \hline
     \large BLIINDS-II \cite{6172573} &   \large 2012 TIP & \large0.4967& \large	0.7172&&  \large 0.2913 
	&  \large 0.6359 
  & & \large	0.4414 &  \large	0.7417 &&  \large	0.5688 	&  \large0.7125 	& & \large0.5810 &  \large	0.7445 
 \\
    \large  BRISQUE \cite{mittal2012no} &  \large    2012 TIP &  \large0.5714 &  \large	0.7306 &&  \large0.3447 
 	&  \large0.6456 
	& &  \large	0.6787 	&  \large0.7838 	&&  \large0.6625 	&  \large0.7813 &&  \large	0.5748 &  \large	0.7305 
  \\
    \large  DEIQT \cite{qin2023data} &  \large    2023 AAAI &  \large 0.6854 	&  \large 0.8098 &&  \large  \textbf{0.4466}
 	&  \large \textbf{0.8155} & &  \large	0.7447 &  \large	0.8559&&  \large 	0.6875 &  \large	0.7438 &&  \large	0.7414 &  \large	0.8271 
 \\
   \large   HyperIQA \cite{su2020blindly} &  \large    2022 CVPR &  \large 0.6496 &  \large	0.7496 &&  \large	0.2379 
	&  \large0.5534 
	&&   \large 	0.7508 &  \large	0.8559& & \large 	0.5938 &  \large	0.6625 	& & \large0.7535 	&  \large0.8019 
  \\
    \large  MANIQA \cite{yang2022maniqa} &  \large   2020 CVPR &  \large 0.6780 &  \large	0.8031 &&  \large	0.4078 
 &  \large	0.5741 		&&  \large0.7136 
 	&  \large0.8318 & & \large	0.5375 	&  \large0.7125 	&  &\large0.7804 &  \large	0.8364 
\\
    \large  Fractal \cite{chen2019effects} &   \large 2019 TGRS &  \large0.6706 &  \large	0.7934& & \large 	0.3816 
 	&  \large0.6748 
	&&  \large0.7310 &  \large	0.8228 & & \large	0.6643 &  \large	0.8625 	&&  \large0.7440 	&  \large0.8114 
  \\
         \hdashline
    \large  Random &  \large  - &  \large 0.1333 &  \large	0.4296 &&  \large	0.1408 
&  \large 0.3301 
  &&  \large	0.1532 &  \large	0.3544&&  \large 	0.1438 	&  \large0.3750 &&  \large	0.1201 &  \large	0.5211 
 \\
         \hdashline
    \large  Ours &  \large    - &  \large \textbf{0.7328} 	&  \large \textbf{0.8253} &&  \large	0.3883 
	&  \large0.6359 
 &&  \large	\textbf{0.7748} &  \large	\textbf{0.8679} & & \large	\textbf{0.8375} 	&  \large \textbf{0.8813}	&&  \large \textbf{0.8069} &  \large	\textbf{0.8645}
 \\
   \hline  \hline
   \end{tabular}
 }
   \end{table*}
\renewcommand\arraystretch{1}
We assess the evaluation capabilities of SQA using Pearson’s linear correlation coefficient (PLCC), Spearman’s rank-order correlation coefficient (SROCC), root mean square error (RMSE) and Kendall rank-order correlation coefficient (KROCC) as metrics. PLCC measures the degree of linear correlation between two sets of variables. SROCC and KROCC calculate the degree of consistency between the ranks of two sets of variables. RMSE is a metric for measuring the difference between predicted values and true values. The smaller the RMSE, the higher the prediction accuracy of the model. The metrics are defined as follows:
\begin{equation}
\text{PLCC} = \frac{\sum_{i=1}^{N} (x_i - \bar{x})(y_i - \bar{y})}{\sqrt{\sum_{i=1}^{N} (x_i - \bar{x})^2 \sum_{i=1}^{N} (y_i - \bar{y})^2}}
\end{equation}
\begin{equation}
\text{SROCC} = 1 - \frac{6 \sum_{i=1}^{N} d_i^2}{N(N^2 - 1)}
\end{equation}
\begin{equation}
\text{RMSE} = \sqrt{\frac{1}{N} \sum_{i=1}^{N} (x_i - y_i)^2}
\end{equation}
\begin{equation}
\text{KROCC} = \frac{2(P - Q)}{N(N-1)}
\end{equation}
where, $x_i, y_i$ represent the predicted and true scores, respectively, for each data point. $\bar{x}, \bar{y}$ represent the mean values of $x_i$ and $y_i$. $d_i$ represents the difference between the ranks of each pair $(x_i, y_i)$. $N$ represents the total number of data points. $P$ and $Q$ represent the number of concordant pairs and discordant pairs, respectively.
Note that PLCC and RMSE are computed after performing a nonlinear four-parametric logistic function to map the objective predictions into the same scale of labels as described in \cite{vqeg_report_2011}.

\subsubsection{Compared Methods}
Open-source unsupervised semantic SQA methods for remote sensing are scarce, making direct comparisons challenging. Therefore, we evaluate RS-SQA against several no-reference image quality assessment (NR-IQA) algorithms. The comparison includes classical metrics such as BLIINDS-II \cite{6172573} and BRISQUE \cite{mittal2012no}, as well as the SOTA deep learning-based methods, including DEIQT \cite{qin2023data}, MANIQA \cite{yang2022maniqa}, and HyperIQA \cite{su2020blindly}. Additionally, we incorporate a no-reference method \cite{chen2019effects} specifically designed for predicting segmentation accuracy in remote sensing images.
We choose BLIINDS-II \cite{6172573} and BRISQUE \cite{mittal2012no} to explore the effectiveness of subjective quality assessment based on the human visual system (HVS). We also select some representative deep-learning NR-IQA methods \cite{yang2022maniqa,su2020blindly,qin2023data} to validate the adaptability of data-driven methods in the task of semantic segmentation quality assessment. The compared methods are retrained on the RS-SQED training set mentioned in Section \ref{section::Image Collection}.

\subsection{Performance and Analysis of RS-SQA on the RS-SQED Dataset}
\label{section::performance}
We trained a quality assessment model for each semantic segmentation method using the RS-SQA framework. To enhance the model's robustness in assessing semantic segmentation quality across diverse scenarios, we employed a mixed dataset for training and conducted comprehensive evaluations on the test set.

The scatter plots between the predicted OA versus their ground truth under eight representative semantic segmentation methods are shown in Fig. \ref {fig-performanceour}. The horizontal and vertical axes denote the predicted OA and the ground truth, respectively, with each data point corresponding to an image.
From the figure, it is observed that there is a strong linear correlation between the predicted performance and the true labels, which is consistent across both the mixed test set and individual semantic segmentation datasets for all seven methods. Additionally, in the mixed test set, the data points are uniformly and densely clustered, suggesting that the proposed method effectively predicts the semantic segmentation accuracy of remote sensing images.
 Among the individual test sets, the UAVid dataset exhibits the best prediction performance, with data points closely aligned along the fitted line. This is likely due to the high ground sampling resolution and precise segmentation annotations, coupled with minimal content differences between the training and test sets of UAVid, which effectively reduce the impact of label uncertainty on the alignment between predicted values and true labels.
Second, the FloodNet dataset exhibits a more pronounced long-tail effect, as most patches belong to one discriminative class. This leads to exceptionally high segmentation accuracy, with many instances achieving 100\%. On the ISPRS dataset, the data points exhibit a significant deviation from the fitted line. This phenomenon can be attributed to the overfitting issue caused by the relatively limited quantity of data within this dataset. Additionally, the disparity in tonal characteristics between this dataset and others renders the semantic features less pronounced, thereby giving rise to inferior performance.
\begin{table*}[htbp]
\centering
\setlength{\tabcolsep}{3pt}
\caption{Ablation Study of the Model Design on the RS-SQED dataset. The precision of Predicting the Best Method among Eight Models. The Bold Indicates the Top One Performer.}
\renewcommand\arraystretch{1.5}
\resizebox{\textwidth}{!}{
\begin{tabular}{c ccccc  cc c cc c cc c cc c cc }
\hline\hline
\multirow{2}{*}{\textbf{Model}}&\multirow{2}{*}{\textbf{Sem. E}}&\multirow{2}{*}{\textbf{Sem. A}} &\multirow{2}{*}{\textbf{Seg. E}}&\multirow{2}{*}{\textbf{Seg.A }}&\multirow{2}{*}{\textbf{SCGB}}& \multicolumn{2}{c}{\textbf{RS-SQED}} && \multicolumn{2}{c}{\textbf{ISPRS}} & &\multicolumn{2}{c}{\textbf{LoveDA}}  & &\multicolumn{2}{c}{\textbf{UAVid}} && \multicolumn{2}{c}{\textbf{FloodNet}} \\ 
\cline{7-8}\cline{10-11}\cline{13-14}\cline{16-17}\cline{19-20}
 && & & &  & \textbf{P@1} & \textbf{P@3}&  &\textbf{P@1} & \textbf{P@3} && \textbf{P@1} & \textbf{P@3}&& \textbf{P@1} & \textbf{P@3}&& \textbf{P@1} & \textbf{P@3}\\ 
\cline{1-20}
w/o adapter &\checkmark & &\checkmark & &\checkmark &0.6304 &	0.7477 &&	0.2796 
 &	0.5720 
 	&&0.6658 	&0.7829 	&&0.6163 &	0.7313 &&	0.7380 	&0.8016 
 \\ 
 w/o SCGB &\checkmark & \checkmark &  \checkmark& \checkmark & &0.6203 	&0.7720 &&	0.1796 
 &	0.5631 
& &	0.6607 	&0.7958 	&&0.7688 &	0.8250 	&&0.7134 &	0.8224 
\\ 
w/o seg. &\checkmark & \checkmark &  & & &0.6721 	&0.7802 &&	0.2282 
 &	0.5437 
 	&&0.7147 &	0.8198 &&	0.6688 &	0.7750 &&	0.8037& 	0.8520 
 \\  
 w/o sem. & && \checkmark& \checkmark& &0.6773 	&0.8061 &&	0.3641 
 	& \textbf{0.6505} 
 &&	0.6607 	&0.7898 	&&0.7500 	&0.8250 &&	0.7788 	&0.8583 
 \\ 
Ours&\checkmark & \checkmark & \checkmark & \checkmark & \checkmark& \textbf{0.7328} 	& \textbf{0.8253} &&	 \textbf{0.3883}
 	&0.6359 
	&&  \textbf{0.7748} &	 \textbf{0.8679} &&	 \textbf{0.8375} & 	 \textbf{0.8813} &&	 \textbf{0.8069}	&  \textbf{0.8645} 
\\ 
\hline\hline
\end{tabular}
\label{tab:ab_model}
}
\end{table*}
\renewcommand\arraystretch{1}
\subsection{Performance Comparison of Different Quality Assessment Metrics}
\label{section::comparion}
To quantitatively assess the effectiveness of our method, we present PLCC, SROCC, RMSE and KROCC results for various comparison methods on the RS-SQED dataset, as tabulated in Table \ref{table::perormanc1}. It is evident that our method significantly enhances the accuracy of semantic segmentation quality assessment for remote sensing images. Compared to existing methods, ours achieves superior performance not only in correlation metrics, such as PLCC and SRCC, but also in terms of RMSE, which is critical for accurately estimating segmentation performance without requiring ground truth labels. Notably, our method maintains consistent performance across various datasets, including the challenging LoveDA dataset, where segmentation accuracy traditionally lags behind due to its complexity.

Traditional image quality assessment methods, namely BLIINDS-II \cite{6172573} and BRISQUE \cite{mittal2012no}, exhibit limited performance in predicting segmentation accuracy, as they are primarily designed to evaluate subjective visual quality in natural images. Deep learning-based methods, such as DEIQT \cite{qin2023data}, perform well on the UAVid and LoveDA datasets, likely due to the large training sets, which enable these methods to capture quality factors related to segmentation accuracy. The poor performance of RS-SQA the ISPRS dataset may be due to the overfitting problem mentioned earlier and the semantic differences with other datasets.

% \subsubsection{Performance on Cross-domain dataset}
\subsection{Application on Recommending the Optimal Semantic Segmentation Method}
\label{section::rank}
To achieve the application-oriented goals that our model not only accurately predict the segmentation accuracy of an image under a specific semantic segmentation method but also assist users in selecting the most accurate method among the numerous remote sensing segmentation options, we calculate the precision of recommending the top-one model among the candidate pool.  

For image $i$, we use the quality assessment model trained with specific OA labels to predict the accuracy for each method, and the predicted scores list is denoted as $S_{i}:\{s_{m1},s_{m2},s_{m3},...,s_{mk}\}$. The method corresponding to the highest score in $S_{i}$ is the semantic segmentation method that our model evaluates as the most suitable for image $i$. The precision is defined as:
\begin{equation}
\text{Precision} = \frac{1}{N} \sum_{i=1}^{N} \mathbb{L}(P_{i}\subseteq B_{i})
\end{equation}
where, $P_{i} $ and $B_{i}$ are the list of predicted best methods and ground truth OA corresponding methods for image $i$. Given that multiple semantic segmentation methods may achieve the same segmentation accuracy for the same image, we define $L$ as an indicator function, indicating that the predicted method list is a proper subset of the true optimal method list.

Specifically, Table \ref{table7} presents the precision of recommending the top-performer across eight semantic segmentation techniques. P@1 denotes the accuracy of identifying the top-ranked model within the model pool, while P@3 represents the accuracy of the selected top-ranked model being among the top three models in the ground truth rankings. Compared to Fractal (the approach most conceptually aligned with ours), our method demonstrates superior performance across all datasets except the ISPRS dataset. The performance gap on ISPRS can be attributed to the poor prediction accuracy of RS-SQA on the semantic segmentation quality score, as shown in Table \ref{table::perormanc1}. This also highlights the difficulty of remote sensing semantic segmentation quality assessment, due to the vast field of view and intricate textures of RSI.
When benchmarked against more sophisticated deep learning-based IQA methods, such as DEIQT and HyperIQA, DEIQT achieves results comparable to ours. However, our method surpasses it by a margin of 5\%, further confirming its efficacy. This finding highlights the capability of RS-SQA to effectively evaluate the unique scene semantics inherent in RSI. With an accuracy of 73\%, RS-SQA reliably recommends the optimal semantic segmentation approach for RSIs.

\begin{figure*}
\centering %表示居中
\includegraphics[height=4.5cm,width=17cm]{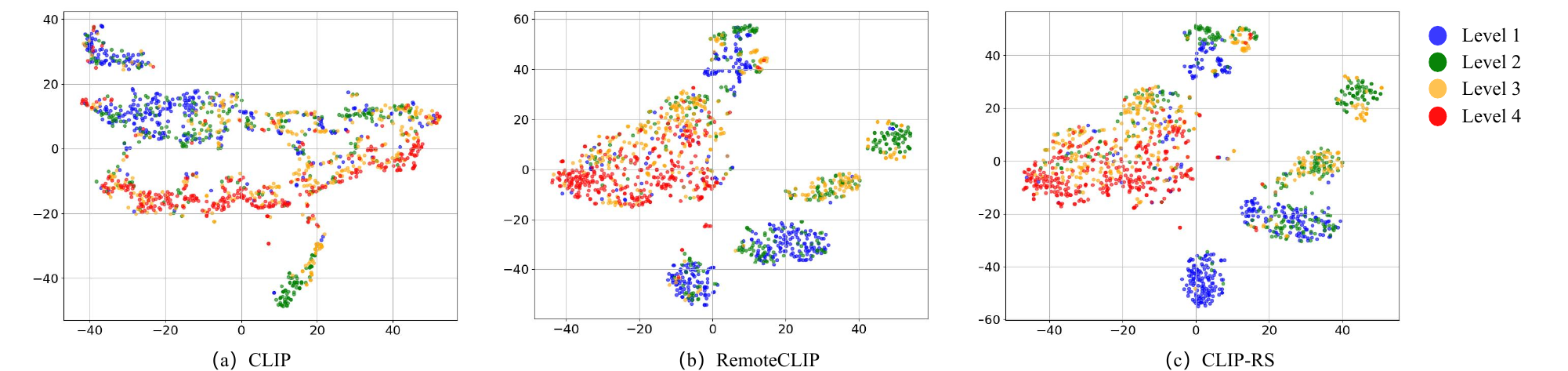}
\caption{ \textbf{t-SNE method visualization of the semantic features extracted from the visual encoder of different VLMs.} The samples are labeled with UperNet((RSP-ViTAEv2-S) segmentation accuracy scores and categorized into four evenly distributed levels.}
%图片的名称
\label{fig-tsne}
%图片的标签，用于文章中的引用，注意到标签的数字与实际文章显示的数字可能不同
\end{figure*}
\subsection{Ablation Study on Recommending the Optimal Semantic Segmentation Method}
In this subsection, we carry out extensive ablation experiments to demonstrate the effectiveness of the model design as well as the large-scale pre-trained VLM in the task of recommending the optimal semantic segmentation method.
\subsubsection{Effectiveness of the Model Design} 
To evaluate the effectiveness of the model design, we conduct a set of experiments on the RS-SQED dataset. We report the Precision defined in Section \ref{section::rank} which reflects not only the correlation but also the prediction accuracy. In Table \ref{tab:ab_model}, ``w/o adapter" indicates that the CLIP-RS visual features are directly used to fuse with the multi-dimensional features extracted by the semantic segmentation method.  These configurations demonstrate the role of the proposed adapter in aligning the two feature types within the latent space. To further validate the effectiveness of SCGB, we replace SCGB with a simple feature concatenation in the ``w/o SCGB" experiment. ``w/o sem." and ``w/o seg." respectively represent predicting by only using features from the segmentation branch or the semantic branch. The results illustrate that the segmentation characteristics encapsulate rich image details and diverse segmentation modalities, and these characteristics play a pivotal role in assessing segmentation quality. Simultaneously, CLIP-RS contributes to a more precise evaluation. Comparative experiments conducted with the ``w/o seg." variant helps to clarify the cause of RS-SQA's unsatisfactory performance on the ISPRS dataset. It is probable that the relatively weak semantic characteristics of the ISPRS dataset are responsible for CLIP-RS's inferior results on this particular dataset.
\subsubsection{Effectiveness of the Loss Function} 
To further validate the effectiveness of the KL divergence joint loss on overall semantic segmentation model evaluation accuracy, additional experiments are conducted. Table \ref{ablationable:Loss} presents the accuracy of predicting the top-performing semantic segmentation method on the RS-SQED dataset across various loss function strategies. The inclusion of $L_{KL}$ contributes to an around 2.6\% gain, indicating the joint loss of $L_{MSE}$ and $L_{KL}$ exhibits substantial improvements in semantic segmentation quality assessment tasks.
\begin{table}[h!]
\caption{Ablation Study of the Loss Function on the RS-SQED dataset. The Precision of Predicting the Best Method among 8 Candidate Methods.}
\label{ablationable:Loss}
\begin{adjustbox}{width=\columnwidth}
\centering
\renewcommand\arraystretch{2}
\begin{tabular}{cccccc}% 其中，tabular是表格内容的环境；c表示centering，即文本格式居中；c的个数代表列的个数
\hline\hline %[2pt]设置线宽     
\large\textbf{Strategy} & \large\textbf{RS-SQED} & \large\textbf{ISPRS}  & \large\textbf{LoveDA} & \large\textbf{UAVid} & \large\textbf{FloodNet}\\ %换行
\hline
\large w/o $L_{MSE}$  & \large0.7195 
 & \large 0.3738 
	 & \large	0.7267 
 & \large0.8125 
	 & \large	0.8146 
 \\ 
\large w/o $L_{KL}$  & \large 0.7061  & \large0.3544 
 	& \large0.7508  & \large	0.7438  & \large	0.7975 \\
\hline
\large  Ours & \large 0.7328 & \large	0.3883 
& \large 	0.7748 	& \large0.8375 & \large	0.8069 
\\
\hline\hline %[2pt]     
\end{tabular}
\end{adjustbox}
\end{table}
\renewcommand\arraystretch{1}

\begin{table}[h!]
\caption{Ablation Study of the Visual Encoder from Different VLMs on the RS-SQED Dataset. The Precision of Predicting the Best Method among 8 Candidate Methods.}
\label{ablationable2}
\begin{adjustbox}{width=\columnwidth}
\label{ablationable}
\centering
\renewcommand\arraystretch{2.3}
\begin{tabular}{ccccccc}% 其中，tabular是表格内容的环境；c表示centering，即文本格式居中；c的个数代表列的个数
\hline\hline %[2pt]设置线宽     
\large\textbf{Model} & \large\textbf{RS-SQED} & \large\textbf{ISPRS}  & \large\textbf{LoveDA} & \large\textbf{UAVid} & \large\textbf{FloodNet}\\ %换行
\hline
\large RemoteCLIP \cite{liu2024remoteclip} & \large 0.6906& \large0.3252 
 & \large	0.7057 & \large	0.6125 & \large	0.8302 
 \\ 
\large CLIP-RS (1.5M)  & \large 0.7276 & \large	0.3839 
  	& \large0.7117 & \large	0.8313 & \large	0.8333 
\\
\hline
\large CLIP-RS (10M)& \large 0.7328 & \large	0.3883 
& \large 	0.7748 	& \large0.8375 & \large	0.8069 
\\
\hline\hline %[2pt]     
\end{tabular}
\end{adjustbox}
\end{table}
\renewcommand\arraystretch{1}
\subsubsection{Effectiveness of the Vision language model}
To verify the effectiveness of large-scale high-quality pre-training, we compare CLIP-RS with RemoteCLIP \cite{liu2024remoteclip}, a representative VLM chosen for its strong performance, on the precision of recommending the best semantic segmentation method. Both CLIP-RS and RemoteCLIP \cite{liu2024remoteclip} are based on the CLIP model, sharing identical structures that allow them to be directly interchangeable without modification to the input and output of other modules. As shown in Table \ref{ablationable}, the CLIP-RS demonstrates significant 4\% advances over RemoteCLIP \cite{liu2024remoteclip}. Notably, the 1.5M variant of CLIP-RS (trained on original high-quality captions, excluding purified captions) also outperforms RemoteCLIP, demonstrating the importance of well-organized datasets in facilitating semantic alignment. On the FloodNet dataset, CLIP-RS (10M) did not consistently outperform the 1.5M variant, possibly due to domain-specific biases, with a larger number of overexposed scenes that may have led the model trained on fewer data to learn domain-specific features. Furthermore, to verify the correlation between semantic information and segmentation accuracy, feature visualizations from the encoders of general CLIP, RemoteCLIP, and CLIP-RS are shown in Fig. \ref{fig-tsne}. These visualization results confirm that CLIP-RS more clearly distinguishes categories related to segmentation quality, thereby enhancing the performance of segmentation quality assessment.

  \section{Conclusion} 
  In this article, we present RS-SQA, a novel semantic segmentation quality assessment framework based on vision language model in the field of remote sensing. The framework employs a dual-branch design, combining high-level semantic features extracted via a big-scale high-quality pre-trained Vision-Language Model, CLIP-RS, with detailed segmentation features to deliver a comprehensive evaluation of segmentation quality.

To support the development of semantic segmentation quality assessment, we establish RS-SQED, a well-crafted dataset covering comprehensive scenarios sampled from 5 commonly used RS semantic segmentation datasets and annotated with accuracy scores for semantic segmentation using 8 different methods.Extensive experiments on RS-SQED demonstrate that RS-SQA outperforms state-of-the-art quality assessment models. The framework is also proved to be highly effective in recommending the most suitable semantic segmentation method, making it a valuable tool for efficient geospatial data processing and analysis to facilitate downstream tasks. Nonetheless, the performance of RS-SQA may be influenced by the diversity and quality of training data. Future work will focus on extending the quality assessment framework to support broader RS tasks.

\bibliography{ms}
% that's all folks
\end{document}